\documentclass[12pt]{article}

\usepackage{tikz}
\usetikzlibrary{positioning, arrows}

\usepackage{amssymb}
\usepackage{amsfonts}
\usepackage{amsmath}
\usepackage{xcolor}
\usepackage{caption}

\numberwithin{equation}{section}

\usepackage{graphicx}

\usepackage[affil-it]{authblk}

\newcommand{\mb}[1]{{\mathbf{#1}}}

\newcommand{\bs}[1]{{\boldsymbol{#1}}}

\def\fddd#1#2{\displaystyle{\frac{\delta #1}{\delta #2}}}

\newcommand{\dsl}[1]{{\displaystyle{#1}}}

\newcommand{\eps}{\epsilon}

\usepackage{graphicx}
\newcommand{\HH}{{{\mathcal{H}}}}
\newtheorem{theorem}{Theorem}[section]
\newtheorem{prop}[theorem]{Proposition}

\newtheorem{remark}[theorem]{Remark}

\newcommand{\ol}[1]{{{\overline{#1}}}}
\newcommand{\wit}[1]{{{\widetilde{#1}}}}
\newcommand{\ou}{{\overline{u}}}

\newcommand{\D}{\mathrm{d}}
\newcommand{\CC}{\mathcal{C}}

\newcommand{\bm}[1]{\mbox{\boldmath{$#1$}}}

\renewcommand{\sfdefault}{bch}
\def\barr{\hbox{{\fontfamily{\sfdefault}\selectfont I\hskip -.35ex R}}}
\def\sbarr{\hbox{{\fontfamily{\sfdefault}\selectfont {\scriptsize I}\hskip -.25ex {\scriptsize R}}}}
\def\ssbarr{\hbox{{\fontfamily{\sfdefault}\selectfont {\tiny I}\hskip -.2ex {\tiny R}}}}
\newcommand{\RR}{{\sbarr}}

\newcommand{\Asf}{{\mathsf A}}
\newcommand{\Bsf}{{\mathsf B}}
\newcommand{\Csf}{{\mathsf C}}

\begin{document}
\title{Hamiltonian reductions, scalings, and\\  effective wave models in stratified fluids  }

\author{Gregorio Falqui ${}^{1,2}$ and Eleonora Sforza  \footnote{Corresponding Author} $ {}^{1,2,3}$}
\affil{
{\small  ${}^1$Department of Mathematics and Applications, 
University of  Milano-Bicocca, \\ Via Roberto Cozzi 55, I-20125 Milano, Italy
}\\ \medskip
{\small  gregorio.falqui@unimib.it, e.sforza4@campus.unimib.it}\\
\medskip
{\small  ${}^2$INFN, Sezione di Milano-Bicocca, Piazza della Scienza 3, 20126 Milano, Italy}\\
\medskip
{\small  ${}^3$Joint Ph.D. program University of Milano - Bicocca, University of Pavia and INdAM}}
\maketitle
\abstract{\noindent We apply 
Poisson reduction techniques 
to describe asymptotic fully nonlinear models of 2-layer sharply stratified fluids in the Hamiltonian framework. 
We start by considering the Benjamin Hamiltonian formalism for a stably stratified $2D$ Euler fluid in a channel of finite height. We use a Marsden--Ratiu reduction scheme for sharply stratified fluids to obtain a canonical formulation of the stratified effective model in one space variable.  
A canonical form of the long-wave Serre-Green Naghdi (SGN) equations is then recovered by means of a suitable double scaling limit in the Hamiltonian function.   Applying the previous results to such a formulation  of the SGN equations, we provide the Miyata Choi-Camassa (MCC) equations for fully non-linear waves in sharply stratified fluids with a natural Hamiltonian structure and discuss the reduced Hamiltonian system obtained taking the natural constraints of the MCC equations into account. To this end, we perform a Dirac-type reduction on a suitable constrained submanifold of fluid field configurations. We finally consider a different scaling limit of the fully non-linear model which leads to a local  Boussinesq-type model in the large-lower layer regime.

\section{Introduction and Outline}
Hamiltonian structures for the Euler equations of water waves and related models have a comparatively long history. In particular, the variational formulation of the Euler equations for an ideal fluid has been the subject of several research efforts, as it presents some challenges with respect to the standard field theory formalism, as possibly first pointed out in \cite{Lin63}.  A decisive breakthrough in this field was obtained by V. Zakharov~\cite{Zak68} with the construction of a canonical structure for the classical Euler water wave equations, later supplemented (see, e.g., ~\cite{Z85,ZK97}) by a Hamiltonian structure for the full Euler equations. 
In this paper we shall reconsider some of these issues in $2$ spatial dimensions, starting from the theory of stratified fluids. 

{
Density stratification in incompressible fluids is an important aspect of theoretical fluid dynamics, and is an inherent component of a wide 
variety of phenomena related to  geophysical application.

Displacement of fluid parcels from their neutral
buoyancy position within a density-stratified flow can result in internal wave motion. The general governing equations of such phenomena are complicated and not easily studied with analytical methods. On the  other hand, dimensionally reduced asymptotic models are often able to effectively describe significant mechanisms of the dynamics.  Over the years many have been proposed in the literature. A partial list includes~\cite{ BB97, Chumaetal09, CGK05,CS93,Du16,PCH, cbj, OVS79,Wu98,Wu2000, CI19} among many others.
}

As mentioned above, a nowadays standard Hamiltonian formulation of the variational problem for the (incompressible) Euler equation is the one set forth by V.E. Zakharov and collaborators. This set-up relies on so-called Clebsch variables for two-dimensional Euler equations

(see, for a review, \cite{ZK97}). 
Our starting point is the Hamiltonian formulation for incompressible flows set up by T.-B. Benjamin \cite{Ben86} (see, also,\cite{BB97}) which involves only physical variables, and is especially powerful in the $2D$ case.

We consider in Section~\ref{SSEF} a sharply stratified two-layer fluid (see Figure \ref{addfig}) confined in a vertical channel.
We consider the  reduction~ \cite{CFO17, CFOPT23} of Benjamin's structure obtained by applying to such an infinite-dimensional system the ideas of the Marsden-Ratiu  (MR) Poisson reduction~\cite{MR86}.    
Our reduced Poisson structure coincides with that devised by Zakharov in his seminal paper~\cite{Zak68}, albeit expressed through "physical" variables.
We asymptotically expand the Euler Hamiltonian in the small dispersion parameter
$\epsilon=h^*/L$, where $h^*$ is a generic vertical length scale and  $L$  a typical wavelenght. In this way,  we compute the total energy of the fluids' system at  $O(\eps^2)$, recovering the results obtained~\cite{CGK05} within the formalism of the {\em Dirichlet-to-Neumann} operator \cite{CS93}.

In Section~\ref{sect: SGN} we specialize $h^*$ to be the asymptotic height $h_2$ of the lower fluid and preliminary get, via a double-scaling limit in which we send the upper fluid's density $\rho_1$ to zero and its asymptotic width $h_1$ to infinity, a canonical version of the celebrated Serre Su-Gardner Green-Naghdi (SGN) equations {(see, e.g.,  \cite{SG69, GN76, LB09, MS85}). Our main results can be described as follows.  We notice that the Miyata  Choi-Camassa (MCC) equations modelling the long-wave expansion of sharply stratified fluids in the so-called hydrostatic approximation can be seen as a pair of SGN equations coupled by the interfacial pressure.} The system encompasses two nontrivial constraints encoding the rigidity of the confining channel and the fluid's volume conservation.
Such geometrical features suggest to
tackle the study of this problem in the Hamiltonian formalism in Section~\ref{SectCC} by the following steps:
\begin{enumerate}
    \item We consider a pair of "canonical" SGN equations as described in Section \ref{sect: SGN}, suitably coupled by the introduction of a interfacial pressure in the momentum balance equations.
    \item We consider the constraints naturally arising in the MCC theory in the light of Dirac's theory of constrained Hamiltonian systems~\cite{Dirac}, and perform the appropriate Dirac-type reduction.
\end{enumerate}
The resulting Hamiltonian system is equivalent to the one described by the MR reduction process performed in Section \ref{evHam}. This somehow closes the circle of ideas and provide the MCC constrained equations a sound Hamiltonian representation, as hinted at in~\cite{CGK05}.

 Finally, as a further application of our scaling recipes,s in Section~\ref{SecDW} we rescale the vertical coordinate  by the asymptotic height 
$h_1$ of the {\em upper} fluid, and we let the width $h_2$ of the lower fluid grow indefinitely. In a suitable asymptotics, we obtain a closed set of Hamiltonian equations which, via a sequence of manipulations,  are shown to be equivalent to a non-purely evolutionary Boussinesq system.

The logical flow of this paper is best described by the {\em flow-chart} in Figure  \ref{flowchart}.
\begin{figure}[h!]
\centering
\includegraphics[width=14cm]{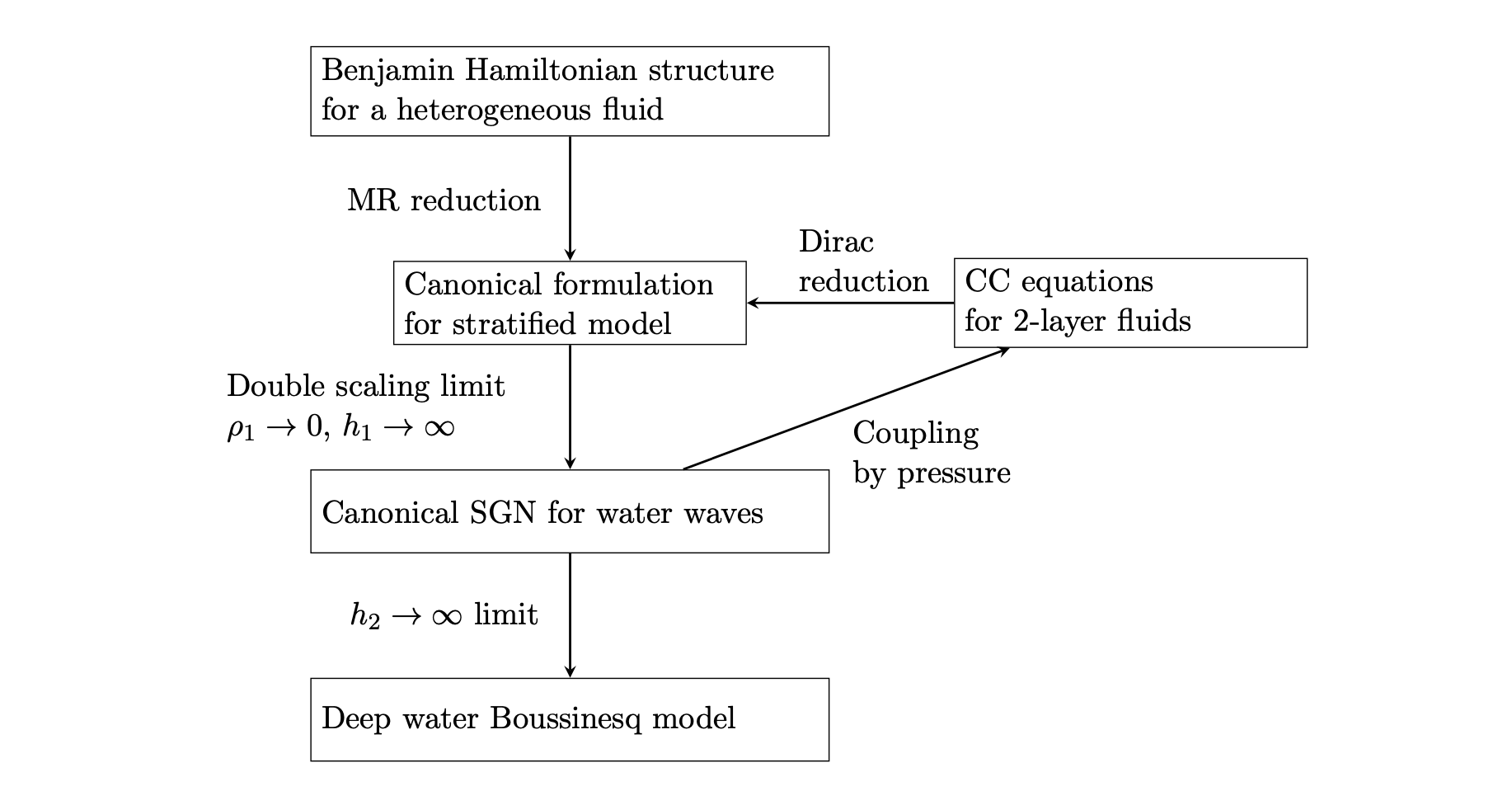}
\captionsetup{width=0.8\textwidth, font=small}
\caption{The flow-chart of the paper}
 \label{flowchart}
\end{figure}

\section{Reduction of the Euler Hamiltonian structure for stratified fluids}\label{SSEF}
 {We consider a perfect, incompressible and variable density fluid confined between two horizontal infinite plates. Thus, the fluid fills a two-dimensional domain $(x,z)\in \barr\times(-h_2,h_1)$, $h\equiv h_1+h_2$ being the distance between the bottom and the upper boundary.
Such a fluid is governed by the incompressible Euler equations for the velocity field $\mathbf{U}=(u,w)$ and non-constant density $\rho(x,z,t)$, in the presence of gravity 
$-g\mathbf{k}$,  
\begin{equation}
\label{EEq}
 {\rho_t}+\mathbf{U}\cdot\nabla\rho=0, \qquad \nabla \cdot \mathbf{U} =0, \qquad \mathbf{U}_t+(\mathbf{U}\cdot\nabla)\, \mathbf{U} + \frac{\nabla p}{\rho} + g \mathbf{k}=0 
 \end{equation}
with boundary conditions 
 \begin{equation}
 \mathbf{U}(x=\pm \infty,z,t)=\mathbf{0},\quad  \text{and } w(x,-h_2,t)=w(x,h_1,t)={0},\quad x\in \barr,
 \quad z\in(-h_2,h_1), \quad t\in \barr^+\,, 
 \label{bEEq}
\end{equation}
where $z=-h_2$ and $z=h_1$ are the locations of the bottom and top confining plates, respectively. }
\subsection{The 2D Benjamin model for heterogeneous fluids in a channel}
\label{Sect-1}The above system was given a Hamiltonian structure in~\cite{Ben86} with basic, locally measurable variables, i.e.,  
the density $\rho$ and  the ``weighted vorticity" $\varsigma$ defined by
\begin{equation}
\label{sigmadef}
{\varsigma}=\nabla\times (\rho\,\bs{U})=(\rho w)_x-(\rho u)_z.
\end{equation}
From~(\ref{EEq}), the equations of motion for these two fields are
\begin{equation}
\label{eqsr}
\left\{
\begin{array}{l}
\rho_t+u\rho_x+w\rho_z =0\\
\varsigma_t+u\varsigma_x +w\varsigma_z +\rho_x\big(gz-\frac12(u^2+w^2)\big)_z+\frac12\rho_z\big(u^2+w^2\big)_x=0\ .
\end{array}
\right.
\end{equation}
These can be written in the form
\begin{equation}
 \label{heq}
{\rho_t}=-\left[\rho,  \dsl{\fddd{\HH}{\varsigma}}\right] \, , \qquad 
\varsigma_t= -\left[\rho,  \dsl{\fddd{\HH}{\rho}}\right]-\left[\varsigma, \dsl{\fddd{\HH}{\varsigma}}\right] \, ,
\end{equation}
where, by definition, $[A, B] = 
A_xB_z-A_zB_x$, and the  functional 
\begin{equation}
\label{ham-ben}
\HH= \dsl{\int_\mathcal{D} \rho\left(\frac12 |\bf{U} |^2+g z\right)\,{\rm d}x\,{\rm d}z}
=\dsl{\int_\mathcal{D} \rho\left(\frac12 |\nabla \Psi|^2+g z\right)\,{\rm d}x\,{\rm d}z}
\end{equation}
is simply given by the sum of the kinetic and potential energy, $\mathcal{D}$ being the fluid domain $\barr\times(-h_2,h_1)$.
The streamfunction $\Psi$ is here used as a placeholder for the map between the weighted vorticity $\varsigma$ and $\bm u$ defined 
by $\varsigma=(\rho \,u)_z-(\rho \,w)_x \equiv -(\rho \, \Psi_z)_z -(\rho \,\Psi_x)_x$.

As shown in~\cite{Ben86}, equations (\ref{heq}) are a Hamiltonian system with respect to a 
linear Lie-algebraic Hamiltonian structure (as it often happens in fluid dynamics, see, e.g., \cite{HMR98, G-BMR12}),  that is,
they can be written as
\[
 \rho_t=\{\rho, \HH\}
,\qquad \varsigma_t=\{\varsigma, \HH\}
\]
for the Poisson brackets defined by the Hamiltonian operator
\begin{equation}\label{B-pb}
P^B=-
\left(\begin{array}{cc}
       0 & \rho_x \partial_z -\rho_z \partial_x \\ 
       \rho_x \partial_z -\rho_z \partial_x & \varsigma_x \partial_z -\varsigma_z \partial_x
      \end{array}
\right).
\end{equation}
Such a setting for incompressible
stratified flows can {\em a posteriori} be seen as the realization of the general Hamiltonian setting via Clebsch gauge variables (see, e.g.,~\cite{Z85}, \cite{MS85} )by means of physical variables as shown in   \cite{CFOP14}.

\subsection{Two-layer case and the Poisson reduction}
\label{2lay}
A very special case of stratified fluid, leading to a notable simplification of the system (\ref{EEq}) that however  retains some of the essential properties of stratification can be obtained by considering a system of two fluids of homogeneous densities $\rho_2>\rho_1$ in the channel $\mathbb R\times (-h_2, h_1)$.
\begin{figure}[ht!]
\centering
\includegraphics[width=13cm]{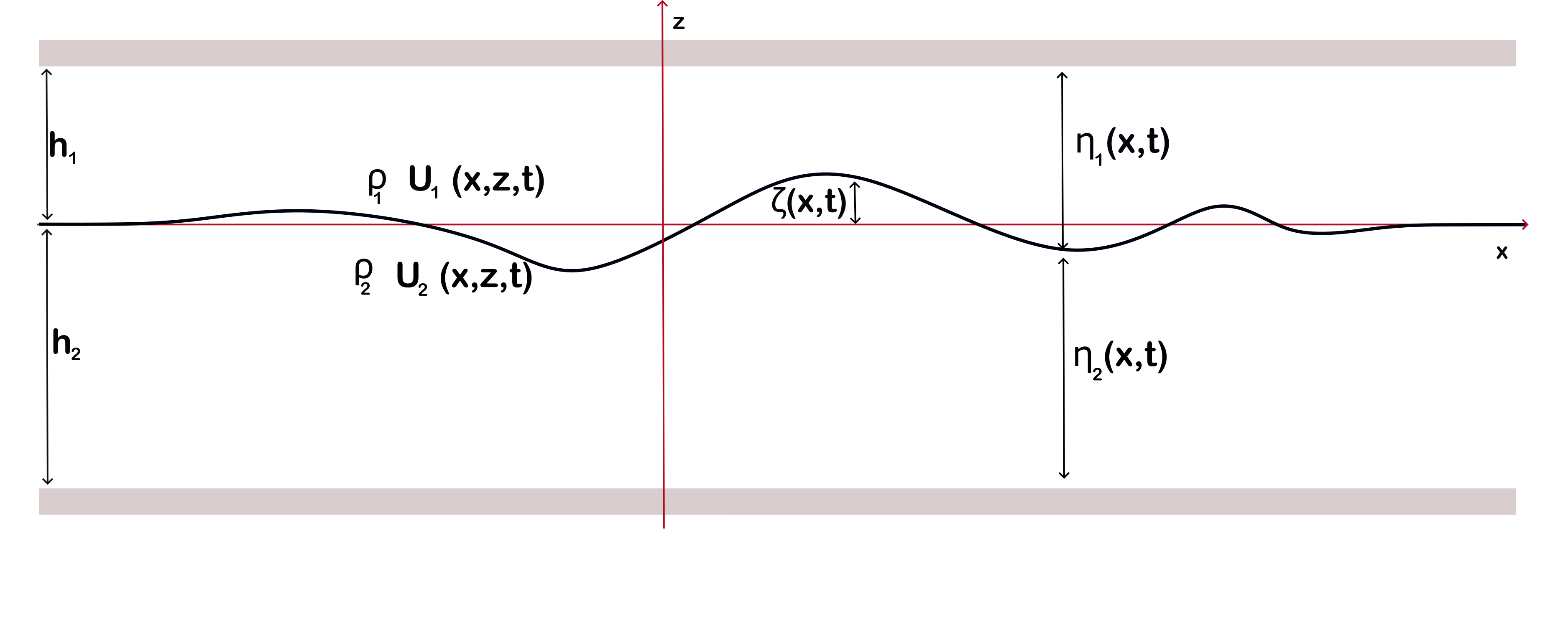}
\captionsetup{width=0.8\textwidth, font=small}
\caption{ Graphical representation of the two-layer configuration. The quantity $\eta_2$ (resp., $\eta_1$) is the total thickness of the lower, heavier (resp., upper, lighter) fluid. The interface  $\zeta$ is measured from the quiescent state $z=0$.
${\bf U}_j(x,z,t)=(u_j(x,z,t), w_j(x,z,t))$ are the velocities vector fields in the two layers. The Hamiltonian variables are related with the interface values, i.e., the  "traces" of the  velocities, $\wit{u}_j(x, t)=u_j(x, \zeta(x,t), t)$, $\wit{w}_j(x,t)=w_j(x, \zeta(x,t), t)$. }
\label{addfig}
\end{figure}

The interface 
between the two homogeneous fluids is described by a smooth function $\zeta=\zeta(x,t)$ { (see Figure~\ref{addfig})}.
In this case the density $\rho$ and the velocity field
 can be described with the aid of the Heaviside function $\theta(\cdot)$  as 
\begin{equation}\label{no1}
\begin{split}
&\rho(x,z,t)=\rho_2+(\rho_1-\rho_2)\theta(z-\zeta(x,t))\\
&u(x,z,t)=u_2(x,z,t)+(u_1(x,z,t)-u_2(x,z,t))\theta(z-\zeta(x,t))\\
&w(x,z,t)=w_2(x,z,t)+(w_1(x,z,t)-w_2(x,z,t))\theta(z-\zeta(x,t))\, ,
\end{split}
\end{equation}
where $\mb{U}_1=(u_1,w_1)$ and $ \mb{U_2}=(u_2,w_2)$ denote the velocity vector fields in the two domains.

{To compute the weighted vorticity (\ref{sigmadef})} we begin with definitions~(\ref{no1}), where we suppress the time dependence for ease of notation in what follows.

The two momentum components are 
\begin{equation}\begin{split} 
&\rho u= \rho_2 u_2(x,z)+(\rho_1 u_1(x,z)-\rho_2 u_2(x,z))\theta(z-\zeta(x))\, , 
\\
&\rho w= \rho_2 w_2(x,z)+(\rho_1 w_1(x,z)-\rho_2 w_2(x,z))\theta(z-\zeta(x)) \, , 
\end{split}
\end{equation}
so that 
\begin{equation}\label{no-s}
\begin{split}
\varsigma=&\rho_2( w_{2\,x}-u_{2\,z})+\big(\rho_1 (w_{1\,x}-u_{1\,z})-\rho_2( w_{2\,x}-u_{2\,z})\theta(z-\zeta(x))
\\&-\big(\rho_1 u_1(x,z)-\rho_2 u_2(x,z)+\zeta_x(\rho_1 w_1(x,z)-\rho_2 w_2(x,z))\big)\delta(z-\zeta(x))\, ,
\end{split}
\end{equation}
where $\delta(\cdot)$ is the Dirac delta function.

We assume that the motion in each layer be irrotational,
so that we are left with a ``momentum vortex line" along the interface, that is, 
\begin{equation}\label{no=s2}
\varsigma=\big(\rho_2 u_2(x,z)-\rho_1 u_1(x,z)+\zeta_x(\rho_2 w_2(x,z)-\rho_1 w_1(x,z))\big)\delta(z-\zeta(x)).
\end{equation}
We define the  projection map 2D $\to$ 1D 
as
\begin{equation}\label{mappazza}
 \zeta(x)=\frac{1}{\rho_2-\rho_1}
 \int_{-h_2}^{h_1} (\rho(x,z)-\rho_1)\, d z \, -h_2, \qquad
{\sigma}(x)=
\int_{-h_2}^{h_1} \varsigma(x,z)\, d z \, .
\end{equation}
When applied to two-layer configurations,  the first of these relations is easily obtained from the first of equations (\ref{no1}).
Moreover, in the two-layer bulk irrotational case, 
\begin{equation}
{\sigma}(x)
=
\rho_2 u_2(x,\zeta(x))-\rho_1 u_1(x,\zeta(x))+\zeta_x(x)(\rho_2 w_2(x,\zeta(x))-\rho_1 w_1(x,\zeta(x))) \, , {\widetilde u}_1(x)+\zeta_x(\rho_2 {\widetilde w}_2(x)-\rho_1 {\widetilde w}_1(x)),
\label{intersi}
\end{equation}
i.e., the averaged weighted vorticity
$%\overline 
\sigma$ is 
the tangential momentum shear at the interface.

{In \cite{CFOPT23} is discussed how this simple averaging process can be given a Hamiltonian structure for stratified homogeneous fluids which fits the Marsden -Ratiu reduction scheme~\cite{MR86}.}

Indeed, such a scheme considers a  manifold $\cal M$ endowed with a Poisson tensor, such as $P^B$,  a submanifold $\mathcal{S}\subset {\cal M}$, a distribution $\mathcal{D}$ contained in the tangent bundle to $\cal M$ restricted to $\mathcal{S}$, $T {\cal M}{\vert_\mathcal{S}}$, and states that a natural Poisson reduction to $\mathcal{S}/\Phi$, with $\Phi$ denoting the intersection $T\mathcal{M}\cap \mathcal{D}$, is possible when (some geometrical assumptions on the regularity of $\mathcal{D}$ and on its action on $\mathcal{M}$ being taken for granted):
\begin{enumerate}
\item $P^B$ is invariant under the distribution $\mathcal{D}$.
\item 
At each point of $\mathcal{M}$ it holds	
\begin{equation}\label{MRcond}
P^B(\mathcal{D}^0)\subset T\mathcal{S}+ \mathcal{D}\, ,
\end{equation}
 \end{enumerate}
$\mathcal{D}^0\subset T^*{\cal M}{\vert_\mathcal{S}}$ being the annihilator of $\mathcal{D}$ in the cotangent bundle to $\cal M$ restricted to $\mathcal{S}$.

In our case, we identify the following geometric objects: 
\begin{enumerate}
\item $\cal M$ is the configuration space  $M^{(2)}$ of 
the 2D fields, parametrized by $(\rho(x,z), \varsigma(x,z))$, and 
$P^B$ is the Benjamin Poisson tensor~(\ref{B-pb})
\begin{equation}\label{B-pbr}
P^B=-
\left(\begin{array}{cc}
       0 & \rho_x \partial_z -\rho_z \partial_x \\ 
       \rho_x \partial_z -\rho_z \partial_x & \varsigma_x \partial_z -\varsigma_z \partial_x
      \end{array}
\right).
\end{equation}
\item  $\mathcal{S}$ is given by the two-layer configuration space
 \begin{equation}\label{intermfld}
\{\rho(x,z)=\rho_2-(\rho_2-\rho_1)
\theta(z-\zeta(x)), \, 
\varsigma(x,z)=
\sigma(x)\delta(z-\zeta(x))\, \}.
\end{equation}
\item $\mathcal{D}$ is the image under $P^B$ of the annihilator $T\mathcal{S}^0$ of the tangent space to $\mathcal{S}$ in $TM^{(2)}\vert_{\mathcal{S}}$.
\end{enumerate} 

{ This choice of the  distribution $\mathcal{D}$  guarantees that  our model fits the Marsden--Ratiu scheme, as it is the field theoretical counterpart of the example ({\bf D}) in \cite{MR86}. Moreover, in \cite{CFOPT23} is shown through direct computations that \begin{equation}
    P^B(\mathcal{TS}^0) = {0}\, , 
\end{equation}
which implies that the reduced manifold $\mathcal{S}/\Phi$ is isomorphic  to $\mathcal{S}$ and the projection \eqref{mappazza} reduces to a change of coordinates. The expression of the reduced  Benjamin Poisson tensor $P^B$ on the manifold ${\mathcal S}=M^{(1)}$   is given, in the coordinates $(\zeta(x), \sigma(x))$, by  the constant  tensor
\begin{equation}\label{Pred}
P^{\rm red}
=-
\left(\begin{array}{cc}
0&\partial_x\\
\partial_x&0\end{array}
\right)\,  .
 \end{equation} }
This structure coincides with the one introduced in~\cite{BB97}  by a direct inspection of the Hamiltonian formulation of two-layer models. We stress that within our setting the above Poisson tensor is obtained by the process of Hamiltonian reduction from the Lie-Poisson structure of the general heterogeneous incompressible Euler $2D$ fluids of \cite{Ben86}. Moreover, by means of our choice of reducing map (\ref{mappazza}), we  directly obtain a set of coordinates $(\zeta,\sigma)$ that can be called {\em Darboux} coordinates, since they are the analog of the 
coordinates $(u,v)$  for the non-linear wave equation in $1+1$ dimensions 
$ u_{tt}=F''(u) u_{xx}$
derived from the Hamiltonian functional ${\mathcal H}=\frac12\int_{\ssbarr} (u_t^2+F(u)) \, \D x $ by means of the Poisson structure (\ref{Pred}). 
\subsection{The Hamiltonian variables { vs. boundary/averaged velocities }}
\label{evHam}
The basic feature of the Hamitonian reduction process is that within this approach the natural dependent variables  are the displacement  from the equilibrium position $\zeta$ 
 and the tangential interface momentum shear
\begin{equation}\label{sigmadef2} \begin{split} 
{{\sigma(x)}}&={\rho_2 u_2(x,\zeta(x))-\rho_1 u_1(x,\zeta(x))+\zeta_x(x)(\rho_2 w_2(x,\zeta(x))-\rho_1 w_1(x,\zeta(x)))}\\ 
& \equiv\rho_2\wit{u}_2(x)-\rho_1\wit{u}_1(x)+\zeta_x(x)(\rho_2\wit{w}_2(x)-\rho_1\wit{w}_1(x))\, \end{split}
\end{equation}
({we recall and use hereafter  that}  
a tilde over a quantity 
{stands} for its evaluation at the interface, e.g., 
$
\wit{u}_1(x,t)=u_1(x, \zeta(x,t), t)$ etc.).

The interface velocity and momentum variables  naturally appear in our Hamiltonian 
approach, while  the 
 Green-Naghdi setting of,  e.g., \cite{CC99},  
 considers {\em layer averaged} velocities (following the seminal paper \cite{Wu81}, see also \cite{Wh2000}).
The link between the two representation described below, can be obtained passing  through another set of coordinates, namely the {\em  boundary} velocities, 
discussed in 
\cite{Wu2000} (see also \cite{BB97, Wh2000, Zak68}).
This goes as follows.

One assumes bulk irrotationality of the fluid flow to introduce the bulk velocity potentials $\varphi_j(x,z)$, which is Taylor-expanded with respect to the vertical variable $z$.
By the vanishing of the vertical velocity at the physical boundaries $z=h_1$, and $z=-h_2$ one gets
\begin{equation}
\label{pot-j}
\varphi_j(x,z)=\sum_{n=0}^\infty \frac{(-1)^n}{(2n)!} 
H_j(z)
^{2n} \partial_x^{2n}\varphi_{0\, j}(x)\, 
\end{equation}
where 
\begin{equation}
\label{Hdef}
H_1(z)=z-h_1, \quad H_2(z)=z+h_2\, ,
\end{equation}
and $\varphi_{0\,1}(x)=\varphi_1(x,h_1)$, $\varphi_{0\, 2}=\varphi_2(x,-h_2)$ are the values of the potential at the rigid lids.
 
The horizontal velocities are then given by
\begin{equation}\label{u-exp1}
u_j=\partial_x\varphi_j(x,z)=\sum_{j=0}^\infty \frac{(-1)^n}{(2n)!} 
H_j(z)
^{2n} \partial_x^{2n}\partial_x\varphi_{0\, j}(x)=\sum_{j=0}^\infty \frac{(-1)^n}{(2n)!} 
H_j(z)
^{2n} \partial_x^{2n} u_{0\, j}(x)\, , 
 \end{equation} 
$u_{0\, j}(x)$ being the horizontal velocities at $z=h_1$ (for $j=1$) and at $z=-h_2$ (for $j=2$).

Likewise, the vertical velocities are given by 
\begin{equation}\label{w-exp1} 
w_j(x,z)=\partial_z\varphi_j(x,z)=\sum_{n=0}^\infty\frac{(-1)^{n+1}}{(2n+1)!}H_j(z)
^{2n+1} \partial_x^{2n+1}u_{0\, j}(x)
\end{equation}
Notice that the boundary conditions $w_1(x,h_1)=w_2(x,-h_2)=0$ are satisfied.
Since \begin{equation}
\label{ex-betaj}
H_1(\zeta)=-\eta_1, \quad H_2(\zeta)=\eta_2,\quad\text{i.e., } H_j(\zeta)=(-1)^j \eta_j,\, j=1,2\, , 
\end{equation}
where $\eta_1(x)=h_1-\zeta(x)$ (resp.\ $\eta_2(x)%=\eta(x)
=h_2+\zeta(x)$) is the thickness of the upper (resp.\ lower) layer,  the interface velocities can be directly obtained by formulas (\ref{u-exp1}) and (\ref{w-exp1}) as
\begin{equation}
\label{intf-speed}
\wit{u}_j=\sum_{j=0}^\infty \frac{(-1)^n}{(2n)!} 
\eta_j^{2n} \partial_x^{2n} u_{0\, j}(x)\, , \qquad \wit{w}_j=(-1)^{j-1}\sum_{n=0}^\infty\frac{(-1)^{n}}{(2n+1)!}\eta_j^{2n+1} \partial_x^{2n+1}u_{0\, j}(x)\, .
\end{equation}
For later use, we express (from the same formulas) the layer-mean horizontal velocities in terms of the fluid thicknesses and the (respective) boundary velocities as
\begin{equation}
\label{mean-vel}
\begin{split}
\ou_1(x)&\equiv \frac{1}{\eta_1}\int_\zeta^{h_1} u_1(x,z)\,\D z=\sum_{n=0}^\infty\frac{(-1)^n}{(2n+1)!}\eta_1(x)^{2n}\partial_x^{2n}u_{0\, 1}(x)\, \\ 
\ou_2(x)&\equiv \frac{1}{\eta_2}\int_{-h_2}^\zeta  u_2(x,z)\,\D z=\sum_{n=0}^\infty\frac{(-1)^n}{(2n+1)!}\eta_2(x)^{2n}\partial_x^{2n}u_{0 \,2}(x)\, .
\end{split}
\end{equation}

\subsection{ Rescaling the spatial independent variables: the $\epsilon$--expansion and the mass conservation laws} \label{rescvar}

To make the formal Taylor series (\ref{intf-speed},\ref{mean-vel}) effective in the construction of asymptotic models for interfacial wave motion we have to rescale variables (see, e.g., \cite{Wh2000, Wu2000}). In particular, we set
\begin{equation}
\label{xz-scale}
x=L\, x^*, \quad z=h^*\, z^*\, , 
\end{equation}
where $L$ is a typical horizontal scale (say, a typical wavelength) and  $h^*$ is a vertical scaling parameter that will later be specialized to assume specific values related with the physical setting of the problem.   As usual, we assume the ratio $\epsilon=\dsl{{h^*}/{L}}$ to be the small dispersion parameter of the theory.
 Making use of such rescalings, we can turn the Taylor series (\ref{u-exp1},\ref{w-exp1}) as well as (\ref{intf-speed},\ref{mean-vel})
into asymptotic series in the small parameter $\eps$.
While this procedure basically follows the approach presented in \cite{CFOPT23}, here we shall not
perform any asymptotic approximations with respect to a small  non-linearity parameter.
For the sake of simplicity, hereafter we shall drop asterisks from adimensionalized formulas. 

In this asymptotics, equations \eqref{intf-speed} and \eqref{mean-vel} yield, respectively,
\begin{equation}
\begin{split}\label{intf-vel-eps}
\wit{u}_j&=
u_{0\, j}-\frac{\eps^2}{2}{\eta_j^2}\, u_{0\,j\, xx}+O(\eps^4)\\
\wit{w}_j&
=\eps\, (-1)^{j-1}\left( 
{\eta_j}{u}_{0\,j\,  x}-\frac{\eps^2
}{6} {\eta_j^3}\,{u}_{0\,j\,  xxx}+O(\eps^4) \right)\, \\ 
\ou_j&=
u_{0\, j}(x)-\frac{\eps^2}{6} {\eta_j^2}\,u_{0\,j\,  xx}+O(\eps^4)\,.
\end{split}
\end{equation}
{It is worth remarking here and below the expected scaling of vertical vs. horizontal velocities $w_j/u_j=O(\eps)$.}
Formulas (\ref{intf-vel-eps})  are obtained taking into account that the quantities $H_j$ and $\eta_j$ scale as the variable $z$, so that 
\begin{flalign}\label{nnn}
    &H_1 = -\frac{h_1}{h^*}+z, \quad H_2 = \frac{h_2}{h^*} + z\\
    &\eta_1(x) = \frac{h_1}{h^*}-\zeta(x), \quad \eta_2(x) = \frac{h_2}{h^*} + \zeta(x). 
\end{flalign}
\color{black}
It should be noticed that contrary to~\cite{Wh2000,Wu2000}, for the time being, we do not rescale the dependent velocity variables $u, w$.

At the leading order in the expansion with respect to the small dispersion parameter $\epsilon$, we have
\begin{equation}
\ol{u}_j=\wit{u}_j, \, \ol{w}_j=\wit{w}_j\simeq 0, \, \text{ with } \sigma=\rho_2\wit{u}_2-\rho_1\wit{u}_1=\rho_2\ol{u}_2-\rho_1\ol{u}_1,
\end{equation}
{that is, $\sigma$ reduces to the {\em horizontal} momentum shear, and interface and layer-averaged quantities coincide.
At higher orders such a property fails, since we have
\begin{equation}
\label{sigmaeps2}
\sigma=\rho_2\wit{u}_2-\rho_1\wit{u}_1+\eps \zeta_x(\rho_2\wit{w}_2-\rho_1\wit{w_1})\, , 
\end{equation}
and, as can be easily obtained from relations \eqref{intf-vel-eps} 
\begin{equation}
\label{utildetou0}
u_{0\, j}=\wit{u}_j+\frac{\eps^2}{2}\eta_j^2\wit{u}_{j\, xx}+O(\epsilon^4)\, ,
\end{equation} 
\begin{equation}\label{utildetowtilde}
\wit{w}_j=(-1)^{j+1}\, \epsilon\, \left( 
%3\eps
\eta_j\wit{u}_{j\, x}+\frac{\eps^2
}{3} (\eta_j \wit{u}_{j\, xx})_x+O(\eps^4)\right)\, , %\, ?? Serve??
\end{equation}
as well as the asymptotic relation between layer-averaged and interface horizontal velocities
\begin{equation}
\label{utildetoou}
\ou_j(x)=\wit{u}_j+\frac{\eps^2}{3}\eta_j^2\wit{u}_{j\, xx}+O(\epsilon^4)\, .
\end{equation}}
The mass conservation laws for the two fluids, {are expressed  by the pair of exact equations}
\begin{equation}
\label{exmce}
\eta_{j\,t}+\partial_x(\eta_j\,\ou_j)=0,\quad j=1,2\, .  
\end{equation}
They are transformed 
by (\ref{utildetoou})  into the approximate mass conservation laws
\begin{equation}
\label{appmce} 
\eta_{j\,t}+\partial_x(\eta_j\,\wit{u}_j) +\frac{\epsilon^2}{3} \partial_x(\eta_j^3\wit{u}_{j \, xx})=O(\eps^4)\,, \quad \, j=1,2\,.
\end{equation}
Hence we can obtain a dynamic constraint  by 
summing the two equations in~(\ref{exmce}) and taking into account the geometric constraint $\eta_1+\eta_2=h$ together with the far-field vanishing conditions, in the form of the
{\em approximate dynamical constraint}
\begin{equation}\label{apprDC}
\eta_1\,\wit{u}_1+\eta_2\,\wit{u}_2+\frac{\epsilon^2}{3}\left(\eta_1^3\wit{u}_{1 \, xx}+\eta_2^3\wit{u}_{2 \, xx}\right)=O(\eps^4)\, .
\end{equation}
\subsection{Asymptotic computation of the reduced energy} 
\label{energ}
Our next task is to write the explicit form (at order $O(\eps^2)$) of the energy in the Darboux coordinates $(\zeta, \sigma)$. 
All the asymptotic manipulations needed are for the kinetic energy,  the potential energy being   
written out immediately.

Let us first consider the lower fluid. Its kinetic energy density reads
\begin{equation}
\label{T2-0} 
T_2=\frac{\rho_2}{2}\, \int_{-h_2/h^*} ^\zeta(u_2^2+ w_2^2)\, h^*\,\D z\, ,
\end{equation}
(the dimensional factor $h^*$ coming from the scaling of {the physical vertical coordinate} $z$).
By Taylor-expanding, we have
\begin{equation} 
u_2(x,z)=u_{2\,0} -\frac{\eps^2}{2} H_2^2u_{2\,0\, xx} +O(\eps^4)\, .
\end{equation}
By (\ref{utildetou0}) we get 
\begin{equation} 
u_2(x,z)=\wit{u}_2+\frac{\eps^2}{2}\left(\eta_2^2-H_2^2\right)\wit{u}_{2\, xx} +O(\eps^4)\, ,
\end{equation}
and by (\ref{intf-vel-eps}), \eqref{nnn}, and (\ref{utildetou0}), 
\begin{equation}
w_2(x,z)=-\epsilon H_2\wit{u}_{2\,x} +O(\eps^3)\, , 
\end{equation}
where the $H_j$'s are defined in~\eqref{nnn}.
This leads to 
\begin{equation}
\label{T2}
\begin{split} 
T_2&=\frac{h^*\, \rho_2}{2}\int_{-h_2/h^*}^\zeta \left[ \wit{u}_2^2+\epsilon^2 \Big( \wit{u}_2\wit{u}_{2\, xx}\big(\eta_2^2-(z+\frac{h_2}{h^*})^2\big)+\wit{u}_{2\,x}^2\, (z+\frac{h_2}{h^*})^2\Big)
+O(\eps^4)\right] \D z \\& =
\frac{h^*\, \rho_2}{2}\left[\eta_2\wit{u}_2^2 +\frac{\eps^2}{3} \eta_2^3 \big(2\wit{u}_2\wit{u}_{2\,xx}+\wit{u}_{2\,x}^2\big) \right]+O(\eps^4) \, .
\end{split}
\end{equation}
By the same arguments we obtain the contribution to the total kinetic energy density of the upper fluid as
\begin{equation}
\label{T1} 
T_1=\frac{h^*\, \rho_1 }{2}\int_\zeta^{h_1/h^*} (u_1^2+ w_1^2)\, \D z=
\frac{h^*\, \rho_1}{2}\left[\eta_1\wit{u}_1^2 +\frac{\eps^2}{3} \eta_{1}^3 \big(2\wit{u}_1\wit{u}_{1\,xx}+\wit{u}_{1\,x}^2\big) \right]+O(\eps^4) \, .
\end{equation}
{In formulas (\ref{T2},\ref{T1}) respectively, the rescaled variables $\eta_1=\dsl{\frac{h_1}{h^*}}-\zeta$ and $\eta_2=\dsl{\frac{h_2}{h^*}}+\zeta$ are involved.}\\
{\color{black} The approximate dynamical constraint \eqref{apprDC} can be rewritten within the present asymptotic theory in the operator form 
\begin{equation}\label{DOpinv}
    \eta_1\left(\mathbf{1}+\frac{\epsilon^2}{3}\eta_1^2\partial_{xx}\right)\wit{u}_{1}=-  \eta_2\left(\mathbf{1}+\frac{\epsilon^2}{3}\eta_2^2\partial_{xx}\right)\wit{u}_{2}\, . 
\end{equation}
Using the inversion formula  
\begin{equation}\label{asinv}
(\mathbf{1} + \epsilon^2 \mathbf{D})^{-1} = \mathbf{1} - \epsilon^2 \mathbf{D}+O(\eps^4)
\end{equation}
 for operators close to identity
 we get, at $O(\eps^2)$,
\begin{equation}\label{u2tildetou1tilde}
\begin{split}
  \wit{u}_1=&-\frac{\eta_2 }{\eta_1}\wit u_2+\frac{\eps^2}{3} \left(\eta_1^2(\frac{\eta_2}{\eta_1}{\wit u}_2)_{xx}-\frac{\eta_2^3}{\eta_1}\wit{u}_{2\,xx}\right) \, , 
  \end{split}
\end{equation}

Combining \eqref{sigmaeps2},  and \eqref{utildetowtilde} we can write the horizontal momentum shear at order $\epsilon^2$ as
\begin{equation}\label{u12tosigma}
    \sigma=\rho_2\wit{u}_2-\rho_1\wit{u}_1-\eps^2 \zeta_x(\rho_2 \eta_2\wit{u}_{2 x}+\rho_1 \eta_1\wit{u}_{1 x})
\end{equation}
which, using \eqref{u2tildetou1tilde},  and introducing  the notation
\begin{equation}\label{lineal}
    \psi = \rho_1 \eta_2 + \rho_2 \eta_1
\end{equation}
for the weighted mass density,  becomes
\begin{equation}\label{u2tosigma}
\begin{split}
       \sigma &= \frac{\psi}{\eta_1} \wit u_2 - \frac{\eps^2}{3}\left(\rho_1 \eta_1^2 (\frac{\eta_2}{\eta_1}\wit{u}_2)_{xx}- \rho_1 \frac{\eta_2^3}{\eta_1}\wit{u}_{2\,xx}\right)-\eps^2\left(\rho_2 \eta_2 \eta_{2\,x} \wit{u}_{2\,x}+\rho_1 \eta_1 \eta_{1\,x} (\frac{\eta_2}{\eta_1}\wit{u}_{2})_{x}\right)
   \, ,
       \end{split}
\end{equation}
where we used $\zeta_x = \eta_{2\,x}= -\eta_{1,x}$.

This gives, by~\eqref{asinv}

\begin{equation}\label{sigmatou2}
       \wit u_2  = \frac{\eta_1 \sigma}{\psi} +\frac{\eps^2}{3 \psi}\left(\rho_1 \eta_1^3(\frac{\eta_2}{\psi}\sigma)_{xx}-\rho_1 \eta_2^3 (\frac{\eta_1}{\psi}\sigma)_{xx}\right)+\frac{\eps^2}{\psi}\left(\rho_2 \eta_1 \eta_2 \eta_{2\,x} (\frac{\eta_1}{\psi}\sigma)_x+\rho_1 \eta_1^2 \eta_{1\,x} (\frac{\eta_2}{\psi}\sigma)_x\right)\,  .
\end{equation}
Then we can substitute this expression in \eqref{u2tildetou1tilde}, to obtain  
\begin{equation}\label{sigmatou1}
    \wit{u}_1=-\frac{\eta_2 \sigma}{\psi} + 
    \frac{\eps^2}{3\psi}\left(\rho_2 \eta_1^3 (\frac{\eta_2}{\psi}\sigma)_{xx}-\rho_2 \eta_2^3  (\frac{\eta_1}{\psi}\sigma)_{xx} \right)-\frac{\eps^2}{\psi}\left(\rho_2 \eta_2 ^2 \eta_{2\,x}(\frac{\eta_1}{\psi}\sigma)_{x}+\rho_1 \eta_1 \eta_2 \eta_{1\,x} (\frac{\eta_2}{\psi}\sigma)_{x}\right)\,.
\end{equation}

Finally, using (\ref{sigmatou2}) and (\ref{sigmatou1}) we arrive at the final formula for the kinetic energy $T=T_1+ T_2$,  given in Darboux coordinates $(\zeta, \sigma)$ by 
\begin{equation}\begin{split}\label{T}
    T &= \frac{h^* \eta_2 \eta_1}{2 \psi}\sigma^2 - \epsilon^2 \, h^* \, \left[
\frac{\eta_2^2 \eta_1^2 \left(\rho_1 \eta_1 + \rho_2 \eta_2\right)}{6 \psi^2} \sigma_x^2  \right. \\[10pt] & \left. +\frac{\rho_1 \rho_2 h\eta_1\eta_2\left(\eta_1^2-\eta_2^2\right)}{6 \psi^3}\zeta_x (\sigma^2)_x  + \frac{\rho_1 \rho_2 h^2\left(\rho_2\eta_1^3+\rho_1 \eta_2^3\right)}{6 \psi^4}\zeta_x^2\sigma^2\right]\, ,
\end{split}
\end{equation}
where we recall that $\psi = \rho_1 \eta_2 +\rho_2 \eta_1$ and $\eta_1=\dsl{\frac{h_1}{h^*}}-\zeta, \eta_2=\dsl{\frac{h_2}{h^*}}+\zeta$, and remark that 
to obtain the comparatively compact formula  \eqref{T} we discarded total $x$-derivatives.

As mentioned at the beginning of this Section, the computation of the potential energy density is more direct.
 Indeed, observing that only variations to the asymptotic equilibrium state 
 \begin{equation}
    \rho_0=\left\{\begin{array}{ll}\medskip \rho_2& -{h_2}/{h^*}\le z<0\\
    \rho_1& 0< z\le h_1/h^*\, , \end{array}
    \right.
\end{equation}
 can enter the energy balance of the system, the potential energy density is

\begin{equation}
\label{Epot} 
U={h^*}^2 g \int_{-h_2/h^*}^{\zeta}\rho z\, dz+\int_{\zeta}^{h_1/h^*} \rho z\, dz - \rho_0= 
 \frac12 {h^*}^2 g (\rho_2-\rho_1) \zeta^2\,,
\end{equation}
where we recall that $h^*$ is the vertical scaling. 

Therefore, the total energy density at this order in the $\eps$ expansion  is
\begin{equation}\label{E}
    \begin{split}
     H  &=  \frac{h^* \eta_2 \eta_1}{2 \psi}\sigma^2 - \epsilon^2 \, h^* \, \left[
\frac{\eta_2^2 \eta_1^2 \left(\rho_1 \eta_1 + \rho_2 \eta_2\right)}{6 \psi^2} \sigma_x^2 +\frac{\rho_1 \rho_2 h\eta_1\eta_2\left(\eta_1^2-\eta_2^2\right)}{6 \psi^3}\zeta_x (\sigma^2)_x \right. \\[10pt] & \left.   + \frac{\rho_1 \rho_2 h^2\left(\rho_2\eta_1^3+\rho_1 \eta_2^3\right)}{6 \psi^4}\zeta_x^2\sigma^2\right]+\frac12 {h^*}^2 g (\rho_2-\rho_1) \zeta^2\, , \\
\end{split}
\end{equation}
where we collect the variables  $\eta_1=\dsl{\frac{h_1}{h^*}}-\zeta, \eta_2=\dsl{\frac{h_2}{h^*}}-\zeta$ and $\psi=\rho_2\eta_1+\rho_1\eta_2$ to improve readability.

The reduced equations of the motion of two stratified fluids can be recovered, at order $O(\epsilon^2)$, applying the reduced Poisson tensor \eqref{Pred} to the variational derivative of the energy $\mathcal{H}= \int_\RR {H\, dx}$, \begin{equation}\label{ham_eq_mot}
    \begin{split}
       \left( \begin{array}{cc}\smallskip
           \zeta_t    \\
            \sigma_t    
            \end{array}\right) = 
            - \left(\begin{array}{cc}\bigskip
0&\partial_x\\
\partial_x&0 \end{array}
\right)\left( \begin{array}{cc}\smallskip
\dfrac{\delta \mathcal{H}}{\delta \zeta} \\
\dfrac{\delta \mathcal{H}}{\delta \sigma}
\end{array}\right)\, .  
    \end{split}
\end{equation}

Clearly enough, the Hamiltonian equations~\eqref{ham_eq_mot} are very complicated, and reporting them not very illuminating. We leave their study for possible  future work. In what follows, 
we shall perform suitable (double) scaling limits to obtain more manageable models. The first of such 
double scaling limits is the ``air-water" one, which will allow us  to recover, within this reduction formalism, the SGN equations, as an intermediate step to study the geometrical derivation of the Miyata  Choi-Camassa equations.

{\bf Remark.} Formula~\eqref{E} for the energy of stratified fluids  (setting $h^*=h_1+h_2$, that is, choosing as a vertical scale the total width of the channel)  coincides with the one of \cite{CGK05} that exploits the theory of  the Dirichlet-Neumann (DN) operators for the water wave  problem.  For the sake of completeness, we briefly discuss such a construction, and its relation with our approach. 

Let $\Omega_i$ be the single fluids' domains, i.e. $\Omega_1= \zeta<z<h_1$, $\Omega_2=-h_2<z<\zeta$ ($x\in \mathbb{R}$) and consider the Laplace problems
\begin{align}
     &\varphi_{i \, xx}+\varphi_{i\, zz}=0\, \quad \text{in}\quad \Omega_i  ,\label{laplacianeq}\\
    &\varphi_{i\,t} + \frac{1}{2}(\varphi_{i \, x}^2+\varphi_{i \, z}^2 )= -\frac{p}{\rho_i}-g z\, \quad\text{in}\quad  \Omega_i\,, \label{berneq}
    \end{align} 
$\varphi_i$ being the velocity potential~\eqref{pot-j},
equipped with  the  boundary conditions
\begin{align}
    &\varphi_{i\,z} = 0\, \quad \text{at}\quad z=(-1)^{i-1} h_i\, , \label{bd_lids}\\
    &\zeta_t = \sqrt{1+\zeta_x^2}\ \, \partial_{n_i}\varphi_i\, \quad \text{at} \quad z=\zeta\, ,\label{bd_interface}\\
  & \varphi_{i\,t} + \frac{1}{2} (\varphi_{i \, x}^2+\varphi_{i \, z}^2)  = -\frac{P}{\rho_i}-g \zeta \, \quad \text{at} \quad z=\zeta\, , 
  \quad \text{at} \quad z=\zeta\, ,\label{bd_cont}
 \end{align}
 where $\partial_{n_i}$ is the  normal  derivative at the interface and $P$ is the interfacial pressure.
The DN operator
maps the Dirichlet data of the Laplace problem to the corresponding Neumann data, namely
\begin{equation}
\partial_{n_i} \varphi_i=G_i[\zeta]\, \wit{\varphi}_i \quad i= 1,2\,,
\end{equation} 
where  we still use the notation $\wit{\varphi}_j$ to denote evauation at the interface, $\wit{\varphi}_i = \varphi_i (x,\zeta)$. 
From~\eqref{bd_interface}, setting  $n\equiv n_2$, 
\begin{equation}
\partial_n \varphi_1 +\partial_n \varphi_2=0\,,
\end{equation}
(where the change in sign accounts for $n_1=-n_2$)
entailing
\begin{equation}\label{Dnconstr}
G_1[\zeta] \wit{\varphi}_1 + G_2[\zeta] \wit{\varphi}_2=0\,. 
\end{equation}

Equation\eqref{Dnconstr} is derived from the kinematic boundary conditions~\eqref{bd_interface}. As such, it is equivalent to 
~\eqref{exmce}, whose sum  yields
\begin{equation}\label{ourcostr}
\partial_x (\eta_1 \bar u_1 +\eta_2 \bar u_2) = 0\,,
\end{equation} that gives  \eqref{apprDC} assuming far-field vanishing conditions.  This connects our setting with the DN one.

\section{The  SGN  Equations}\label{sect: SGN}

The  water wave setting, whose geometry  is depicted in Figure \ref{Fig2},  is   is recovered by the two-fluid configuration  by means of 
\begin{description}
\item[ i)]  The choice $h^*=h_2$ for the vertical lenght scale;
\item[ii)]  the (double) scaling limit 
\begin{equation}
\label{limit}
\rho_1\to 0, \quad h_1\to\infty,
\quad\text{with }\quad  
\rho_1h_1\to 0\, .
\end{equation}
\end{description}
The Hamilton structure is invariant under these choices and, in particular, under the scaling limit ii).
 The energy density,  obtained from \eqref{E} reduces to
\begin{equation}
   H = \frac{1}{2} \left[\frac{h_2 \eta_2}{\rho_2}\sigma^2 - \epsilon^2 \, \frac{h_2\eta_2^3}{3 \rho_2}\sigma_x^2+ h_2^2 g \rho_2 \zeta^2\right]\, , 
\end{equation}
where  the asymptotic relation
\begin{equation}
        \frac{\eta_1}{\psi} \rightarrow \frac{1}{\rho_2}\, 
    \end{equation}
    ensuing from \eqref{limit},  has been extensively used.

The energy of the water wave ``limit" is already in the Darboux conjugated variables since now  $\eta_2$ is related to $\zeta$ by the trivial translation $\eta_2=1+\zeta$. 
    \begin{figure}[ht!]
\centering
\includegraphics[width=13cm]{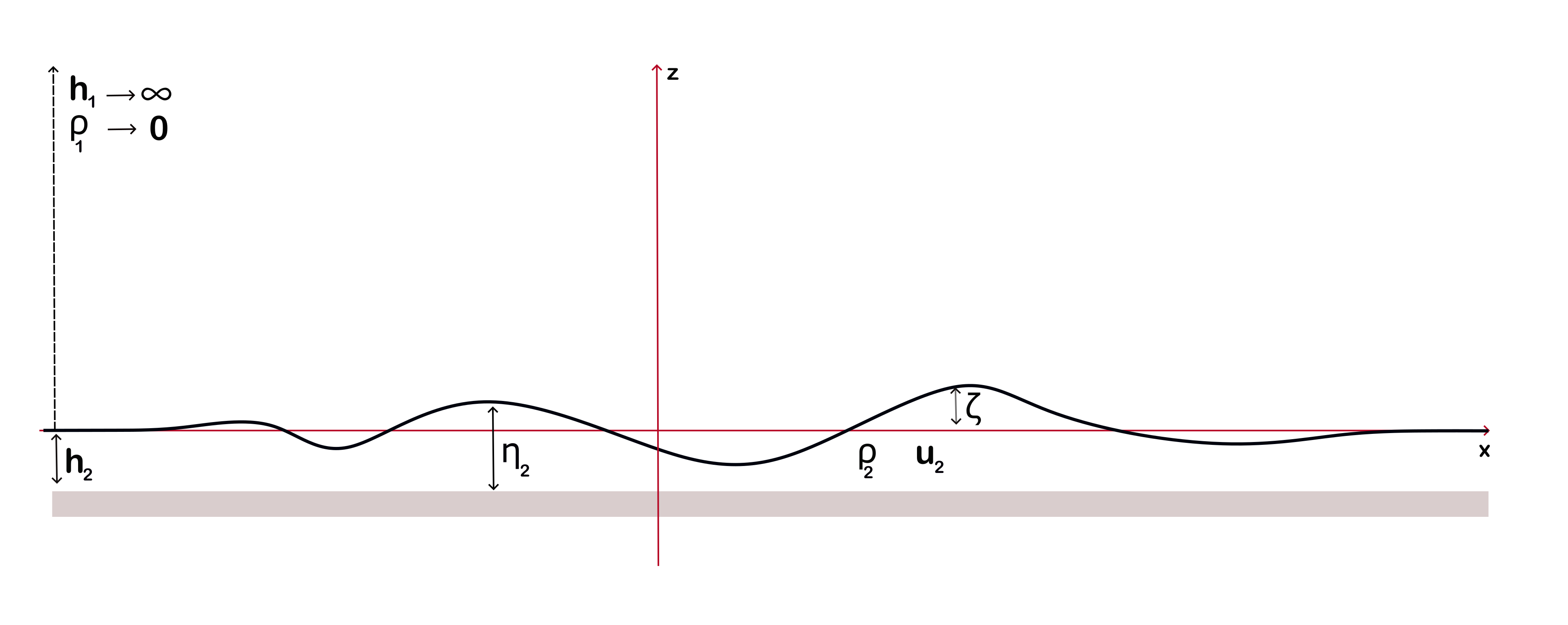}
\captionsetup{width=0.9\textwidth, font=small}
\caption{ The Air-Water limit. { The vertical scaling parameter is the asymptotic height $h_2$ of the lower fluid. The double scaling limit  $h_1=\infty$, $\rho_1\to 0$ with the product $\rho_1 h_1\to 0$  is performed. As it will be shown in this Section (and can naturally be expected), the equations of the stratified model reduce to the Serre Green-Naghdi fully non-linear water wave equations} }
\label{Fig2}
\end{figure}
In the rest of this Section, we shall drop the subscripts referring to the lower fluid and denote $\wit{u}_2=\wit{u}, \eta_2=\eta, h_2=h, \rho_2=\rho\, .$  Also, for reasons that will be apparent in the sequel,  in this single layer effective model we prefer to call $\mu$ the canonical variable $\sigma$. With these conventions relation \eqref{u12tosigma} reads
\begin{equation}\label{Hvars}
\mu=\rho\left(\wit{u}-\eps^2 \eta\eta_x\wit{u}_x\right)\, ,\end{equation} and the 
 Hamiltonian functional reads, after the rescaling $\mu\to \dsl{\frac{\mu}{\sqrt{h}}}$ 
\begin{equation}
\label{Haw}
\mathcal{H}[\eta,\mu]=\frac{h^2} 2\int_\RR \left(\frac{1}{\rho}(\eta\mu^2-\frac{\eps^2}{3}\eta^3{\mu_x} ^2)+g \rho(\eta-h)^2\right)\, d\, x\, .
\end{equation}
The ensuing Hamiltonian equations of motion obtained as
\begin{equation}
    \label{wwham}
\left\{
\begin{array}{l}\medskip
\dsl{\eta_t+\partial_x(\frac{\delta \mathcal{H}}{\delta \mu})=0}\\
\dsl{\mu_t+\partial_x(\frac{\delta \mathcal{H}}{\delta \eta})=0}\, ,
\end{array}
\right.
\end{equation}}
are, via a suitable time rescaling 
\begin{equation}
\label{Heqm}
\left\{
\begin{array}{l}\medskip
\dsl{\eta_t+\frac{(\eta\mu)_x}{\rho}+\frac{\eps^2 }{3\, \rho}(\eta^3\mu_x)_{xx}}=0\,  \\
\dsl{\mu_t+\frac{\mu\mu_x}{\rho}+ g \rho\eta_x-\frac{\eps^2 }{2\, \rho}(\eta^2{\mu_x}^2)_{x}}=0\, .
\end{array}
\right.
\end{equation} 

\begin{prop}\label{Proaw}
{\rm
Equations (\ref{Heqm}) are equivalent to the 2-dimensional  Serre Green-Naghdi (SGN) system (see, e.g., \cite{Ser56,SG69, CHH94, Ma15})
\begin{equation}
\label{SGNeq}
\left\{
\begin{array}{l}\medskip
\dsl{\eta_t+(\eta\ou)_x}=0\,  \\
\dsl{\ou_t+\ou\,\ou_x+g \eta_x-\frac{\eps^2}{3\, \eta}\left(\eta^3(\ou_{tx}+\ou\,\ou_{xx}-\ou_x^2)  \right)_x=0}\, ,
\end{array}
\right.\,  
\end{equation}
where the SGN variable $\ou$ is the layer averaged velocity
\begin{equation}
\label{av vel}
\dsl{ \ou(x)=\frac1{\eta(x)}\int_{-h}^{\zeta(x)} u(x,z)\, d\, z\, .}
\end{equation}
}
\end{prop}
{\bf Proof.}
The key fact is the $O(\eps^2)$ relation between the Hamiltonian variable $\mu$ and the SGN variable $\ou $, that reads
\begin{equation}
\label {outomu}
\mu(\eta, \ou)=\rho( \ou-\dsl{\frac{\eps^2}{3\eta} (\eta^3\ou_x)_x})\, ,
\end{equation}
which follows from \eqref{Hvars}  thanks to the  asymptotic relations
\begin{equation}
\ou=\wit{u}+\frac{\eps^2}{3}\eta^2{\wit{u}}_{xx}\, \Leftrightarrow\,  \wit{u}=\ou-\frac{\eps^2}{3}\eta^2{\ou}_{xx}\, .
\end{equation}
{The conserved energy (\ref{Haw}) rescaled by $\dsl{\frac1{h^2}}$ and written in $(\eta,\ou)$ coordinates reads, after integrating by parts, 
\begin{equation}
\label{Esgn}
\mathcal{H}[\eta,\ou]=\frac12 \int_\RR \rho (\eta\ou^2+\frac{\eps^2}{3}\eta^3\ou_x^2+g(\eta-h)^2)\, d\, x\, ,
\end{equation}
and coincides with the energy of the SGN system (see, e.g.,  \cite{CHH94, Ma15}). 

One notices that the conjugate momentum of the  SGN theory, 
\begin{equation}
\label{mdef}
m=\dsl{\frac{\delta }{\delta \ou}H[\eta,\ou]=\rho(\eta\ou-\frac{\eps^2}{3}(\eta^3\ou_x)_x)}
\end{equation}
coincides, at $O(\eps^2)$, with the product $\eta\mu$ of our Darboux variables. Finally, under the coordinates transformation
\begin{equation}
\label{drtosgn}
(\eta,\mu)\to (\eta,m=\eta\mu)\, ,
\end{equation}
the Darboux Poisson tensor
\begin{equation}
\label{P-D}
P^{\rm red} =-\left(
\begin{array}{cc} 0&\partial_x\\
\partial_x&0\end{array}
\right)
\end{equation}
transforms into the Lie-theoretic linear  Poisson tensor
\begin{equation}
\label{P-sgn}
P^{SGN}=-\left(
\begin{array}{cc} 0&\partial_x\eta\\
\eta\partial_x&m\partial_x+\partial_x m\end{array}
\right)\, ,
\end{equation}}
of the SGN theory.
\hfill$\blacksquare$\\

{\bf Remark.}
This proposition, that recovers the results of ~\cite{Ma15}, could also be proven by a straightforward but tedious  computation. 
Admittedly, the fact that the double scaling limit $\rho_1\to 0, h_1\to\infty$ yields the SGN equations for air-water systems is not surprising at all. However, we stress that our derivation is geometrical and shows how these equations can be derived in a Hamiltonian reduction process from the corresponding Hamiltonian formulation of the Euler equations. This representation will be instrumental to provide the $2$-layer Miyata Choi-Camassa (MCC) equation with a canonical Hamiltonian structure in the next Section.

\section{A Hamiltonian representation of the Miyata Choi-Camassa  equations and its reduction}\label{SectCC}
The  Miyata Choi-Camassa (MCC) equations \cite{CC99, M88} describe the dynamics of a two-layered sharply stratified $2$-dimensional fluid confined 
in a channel of fixed height  $h=h_1+h_2$,  $h_1$ (resp. $h_2$) being the far-field height of the upper (resp. lower) fluid, with densities $\rho_1<\rho_2$  (see the discussion and Figure \eqref{addfig} at the beginning of  Section \ref{2lay}). 
They were deduced via a careful analysis of the long wave asymptotic expansion of the Euler equations and read
\begin{equation}
\label{MCCeq}
\left\{
\begin{array}{l}
\dsl{\eta_{k\, t}+(\eta_k\ou_k)_x}=0\,   \\ \medskip
\dsl{\ou_{k\,t}+\ou_k\ou_{k\, x} {+ g \eta_{k\, x}}-\frac{\eps^2}{3\, \eta_k}\left(\eta_k^3(\ou_{k\, t x}+\ou_k\ou_{k\, xx}-\ou_{k\, x}^2)  \right)_x=-\frac{P_x}{\rho_k}}, \\
k=1,2,\quad\text{with } \eta_k=h_k+(-1)^k\zeta\, ,
\end{array}
\right.
\end{equation}
where $P=P(x,t)$ is the (unknown) interfacial pressure, and $\ou_k$ are the layer-averaged velocities
\begin{equation}
\label{Wu-av}
\dsl{\ou_1=\frac1{\eta_1}\int_{\zeta}^{h_1} u(x,z)\, d\, z\, , \qquad  \ou_2=\frac1{\eta_2}\int_{-h_2}^{\zeta} u(x,z)\, d\, z\, .}
\end{equation}
They clearly can be seen as a pair of SGN equations (the one of the upper layer $k=1$ with reversed gravity), coupled by an "external" field $P(x,t)$. 
They come equipped with the constraints 
\begin{equation}
\label{dyn-con}
\eta_1+\eta_2=h\, , \quad
\eta_1\ou_1+\eta_2\ou_2=0\, , 
\end{equation}
the latter "dynamical" constraint deduced by  adding the two mass conservation equations $\eta_{k\, t}+(\eta_k\ou_k)_x=0\,,  k=1,2$ and taking into account  that $\eta_1+\eta_2=h$  and the far-field vanishing conditions.

Taking into account the results of Section \ref{SectCC}, the MCC equations acquire a natural Hamiltonian formulation by replacing the upper and lower averaged densities with a pair of momentum variables $\mu_k, k=1,2$ via the relations
 \begin{equation}
\label{momenta}
\mu_k(\eta_k, \ou_k)=\rho_k( \ou_k-\dsl{\frac{\eps^2}{3\eta_k} (\eta_k^3\ou_{k\, x})_x}\, , \quad k=1,2\, .
\end{equation}
Considering  the product Darboux Poisson structure
\begin{equation}
\label{P4}
P^{(4)}=-\left(\begin{array}{cccc} 0&\partial_x&0&0\\
\partial_x&0&0&0\\ 0&0&0&\partial_x\\ 0&0&\partial_x&0\end{array}
\right)
\end{equation}
and the Hamiltonian functional  
\begin{equation}
\label{HMCC}
\begin{split}
H_{CC}[\eta_1,\mu_1,\eta_2,\mu_2] 
&=\frac12\int_\RR \left(\frac1{\rho_1}(\eta_1\mu_1^2-\frac{\eps^2}{3}\eta_1^3\mu_{1\, x}^2)-g\rho_1(\eta_1-h_1)^2 +\eta_1\, P)\right)\, d\, x\, +
\\ &\frac12\int_\RR \left(\frac1\rho_2(\eta_2\mu_2^2-\frac{\eps^2}{3}\eta_2^3\mu_{2\, x}^2)+g\rho_2(\eta_2-h_2)^2 +\eta_2\, P\right)\, d\, x\,,\end{split}
\end{equation}
we see that a canonical Hamiltonian form for the MCC equations \eqref{MCCeq} is obtained by setting
\begin{equation}
\label{HMCCformal}
\left(\begin{array}{c}
\eta_{1\, t}\\
\mu_{1\, t}\\
\eta_{2\, t}\\
\mu_{2\, t}\end{array}
\right)=-\left(\begin{array}{cccc} 0&\partial_x&0&0\\
\partial_x&0&0&0\\ 0&0&0&\partial_x\\ 0&0&\partial_x&0\end{array}
\right)\cdot \left(\begin{array}{c}
\delta_{\eta_{1}} H_{MCC}\\
\delta_{\mu_{1}}H_{MCC}\\
\delta_{\eta_{2}} H_{MCC}\\
\delta_{\mu_{2}} H_{MCC}\end{array}
\right)\, , 
\end{equation}
that is, explicitly, 
\begin{equation}
\label{HMCCexpl}
\left\{
\begin{array}{l}
\dsl{\eta_{k\, t}+\frac{(\eta_k\mu_{k})_x}{\rho_k}+\frac{\eps^2}{3\, \rho_k}(\eta_k^3\mu_{k\, x})_{xx}}=0\,  \\
\dsl{\mu_{k\, t}+\frac{\mu_k\mu_{k\, x}}{\rho_k}+{ g\rho_k\eta_{k\, x}}-\frac{\eps^2}{2\, \rho_k}(\eta_k^2 \mu_{k\, x}^2)_{xx}}+P_x=0\, ,
\end{array}
\right.
\end{equation}
with $k=1,2$.
The counterparts of the constraints~\eqref{dyn-con} read, in these coordinates and at $O(\eps^2)$ 
\begin{equation}
\label{muetacon}
\eta_1+\eta_2=h\, , \qquad  \dsl{\sum_{k=1}^2\frac1{\rho_k}\left( \eta_k\mu_k+\frac{\eps^2}{3}(\eta_k^3\mu_{k\, x})_x\right)=0}\, .
\end{equation}

{\bf Remark.} The full  MCC system can be seen as {\em constrained} system, the primary constraint being the geometrical one $\eta_1+\eta_2=h$, with the interfacial pressure $P$ playing the role of Lagrange multiplier in the energy. 
The coordinate change to our Darboux coordinates has the advantage of setting the equations in a simple Poisson form, at the price of losing the exactness of the conservation laws for the height variables $\eta_i$.  Also, the MCC model describes the same physical problem with the same asymptotics (shallow water) as in Section \ref{2lay} (see also Figure \ref{addfig}). For this reason, one can expect that the Dirac-reduced form of these equations coincides with \eqref{ham_eq_mot}. This is what we show in the arguments to follow.

\subsection{The Dirac-reduced Poisson operator}\label{DREq}

We herewith discuss how the "canonical" form of the MCC equations (\ref{HMCCexpl}) can be reduced to the  constrained manifold $\mathcal {C}$ defined by  (\ref{muetacon}) to a closed system of two Hamiltonian equations in the  two free variables $\zeta=\dsl{1/2\, (\eta_2-\eta_1)}$ and $\sigma=\mu_2-\mu_1$. To this end we shall use the classical Dirac theory of constraints \cite{Dirac}.

Let us denote by $\phi_1=0$ the first ("geometrical ") constraint $\eta_1+\eta_2-h=0$, and notice that the second constraint in (\ref{muetacon}), to be hereafter written as $\phi_2=0$,  is the Dirac secondary constraint obtained by time evolution  of $\phi_1$. Actually, the system of constraints $\{\phi_1=0,\phi_2=0\}$ is complete in Dirac's sense, that is,  at $O(\eps^2)$ it holds:
\begin{prop}
The time evolution of $\phi_2$ vanishes on the  manifold  defined by $\{\phi_1=0,\phi_2=0\}$, i.e., in the notations of Dirac's theory, 
\begin{equation}
\phi_{2\,t}\approx 0
\label{phi2t}
\end{equation}
\end{prop}
{\bf Proof.} The key property quickly leading to the proof of (\ref{phi2t}) is that $\phi_{2,x}$  is the time-derivative of the geometrical constraint $\phi_1=\eta_1+\eta_2-h=0$. Hence, 
$(\phi_{2 \, t})_x=(\phi_{2 \, x})_t=\phi_{1\, tt}=0$, i.e., $\phi_{2,t}$ is independent of $x$, and thus can be computed in the limit $|x|\to\infty$.

This observation and the time evolution equations (\ref{HMCCexpl})  yield
\begin{equation}
\label{phi2tcomp}
\dsl{\phi_{2\, t}\equiv\phi_{2\, t}^\infty=-\left(\frac{h_1}{\rho_1}+\frac{h_2}{\rho_2}\right) P^\infty_x-\frac{\eps^2}{3}\left(\frac{h^3_1}{\rho_1}+\frac{h^3_2}{\rho_2}\right) P^\infty_{xxx}}\, ,
\end{equation}
the superscript ${}^\infty$ standing for the $x\to\infty$ values of the corresponding quantities. So the time-derivative $\phi_{2\, t}$ vanishes since in the limit of large 
$|x|$ the pressures $p_k$ of the fluid layers must equal their hydrostatic limits, i.e.
\begin{equation}
\label{Pinfinity}
\lim_{x\to \pm\infty} p_k(x,z)=-\rho_k g z+P^\infty_\pm\, 
\end{equation}
with $P^\infty_\pm$ two (possibly different, see \cite{Ben86, CCFOP12}) constants.\hfill $\blacksquare$\medskip\\
{Once a complete set of constraints has been found,} to explicitly perform the reduction we  have to compute the Dirac-reduced Poisson tensor, and  determine the restricted Hamiltonian
$H_{MCC}\big\vert_{ \phi_1=0, \phi_2=0}$ from (\ref{HMCC}). {

On general grounds, given an  $M$-dimensional  Poisson manifold $(\mathcal{M},  \{\cdot, \cdot\})$ and a submanifold $\mathcal{S}$  defined by the vanishing of $N$ independent constraints $\phi_a=0, a=1,\ldots, N$,  under the condition that the matrix $\mathsf{C}_{ab}=\{\phi_a,\phi_b\} $  be invertible the Dirac brackets  \cite{Dirac}  
are defined by
\begin{equation}
\label{DirFD}
\{g, f\}^D=\{g,f \}-\sum_{a,b=1}^N \{g, \phi_a\}(\mathsf{C}^{-1})_{ab} \{\phi_b, f\}\, , 
\end{equation} 
and they induce a corresponding Poisson structure (called Dirac-reduced Poisson structure) on $\mathcal{S}$.
A suitable  procedure to compute the associated Dirac reduced Poisson tensor goes as follows:
 \begin{enumerate}
\item Pick a set of coordinates adapted to the constraints, that is, consider the coordinate set $(y_{j_1}, \ldots, y_{_{M-N}}, \phi_1, \ldots, \phi_N)$, that is complement the constraints with independent coordinates $y_k, k=1,\ldots, M-N$.

\item Write the Poisson tensor $P$ corresponding to the original bracket $\{\cdot, \cdot\}$ in these new coordinates to obtain the new matrix representation
\begin{equation}
\label{PDtilde}
\wit{P}=\left( \begin{tabular}{c|c} $\Asf$&$\Bsf^T$
\\ \hline 
 $-\Bsf$&$\Csf$\smallskip\end{tabular}
\right)\, \,.
\end{equation}
\item Observe that the representation of the Dirac bracket~\eqref{DirFD} is given by
\begin{equation}
\label{DirRP}
\wit{P}^D=\left(
\begin{array}{cc}
\mathsf{A}- \mathsf{B}\cdot\mathsf{C}^{-1}\cdot \mathsf{B}^T&0\\0&0
\end{array}\right)\, , 
\end{equation}
and in particular, that the constraint functions  $\phi_a$ are Casimir functions of~\eqref{DirFD}.
\item Finally, naturally obtain the Dirac reduced Poisson tensor in the free coordinates $(y_1, \ldots, y_{M-N})$ on $\mathcal{S}$ 
 by the $(M-N)\times(M-N)$  north-west block of \eqref{DirRP}, i.e., 
\begin{equation} \label{pdr}
\wit{P}^D_r=\Asf- \Bsf\cdot\Csf^{-1}\cdot \Bsf^T\, .
\end{equation}
\end{enumerate}
The Dirac brackets~\eqref{DirFD} were  originally  introduced for finite dimensional Hamiltonian systems.  However, the procedure to obtain the reduced tensor can be  transplanted to field-theoretic models, such as the MCC system and other models (see, e.g., \cite{CFOPT23} and \cite{CFOPS26}).

To apply  the procedure herewith briefly described both to compute the reduced tensor and the constrained Hamiltonian, we consider the change of variables
\begin{equation}
\label{chv}
(\eta_1,\mu_1,\eta_2,\mu_2)\to(\zeta, \sigma,\phi_1,\phi_2)\, ,
\end{equation}
where $\zeta$ is the displacement from the stationary widths $\eta_k=h_k$, $\sigma$ is the tangential momentum shear, while $\phi_1,\phi_2$ are the constraints.
Such coordinate  change  explicitly reads 
\begin{equation}
\label{expchv}
\left\{
\begin{array}{l}
\medskip
\zeta=\dsl{\frac12\,(\eta_2-\eta_1)},\quad
\sigma=\mu_2-\mu_1,\\ \medskip
\phi_1=\eta_1+\eta_2-h,\\
\phi_2=\dsl{\sum_{k=1}^2\frac1{\rho_k}( \eta_k\mu_k+\frac{\eps^2}{3}(\eta_k^3\mu_{k\, x})_x)}\, .
\end{array}
\right.
\end{equation}
The Poisson tensor $P^{(4)}$ of Eq. (\ref{P4}) is written in this new set of coordinates as 
\begin{equation}
\label{Pnewg}
\widetilde{P}^{(4)}=J\cdot P^{(4)}\cdot J^T\, , 
\end{equation}
$J$ being the (Fr\'echet) Jacobian of the transformation (\ref{expchv}) 
\begin{equation}
\label{Jacob}
J=\left(
\begin{array}{cccc}\medskip
\dsl{-\frac12}&0&\dsl{\frac12}&0\\
0&-1&0&1\\
1&0&1&0\\
\delta_{\eta_1}\phi_2&\delta_{\mu_1}\phi_2&\delta_{\eta_2}\phi_2&\delta_{\mu_2}\phi_2
\end{array}
\right)
\end{equation}
where the last line is the Fr\'echet differential of the constraint $\phi_2$ w.r.t. the new coordinates, and matrix multiplication of operators implies composition of operators.

A  direct computation shows that 
the transformed Poisson tensor $\widetilde{P}^{(4)}$ has the form
\begin{equation}
\label{Ptrasf}
\widetilde{P}^{(4)}=\left(
\begin{array}{cccc}
0&-\partial_x&0&\widetilde{P}_{1 4}\\
-\partial_x&0&0&\widetilde{P}_{2 4}\\
0&0&0&\widetilde{P}_{3 4}\\
-\widetilde{P}_{1 4}^T&-\widetilde{P}_{2 4}^T&-\widetilde{P}_{3 4}^T&\widetilde{P}_{4 4}
\end{array}
\right)\, ,
\end{equation}
for suitable differential operators $\widetilde{P}_{i 4}, i=1,\ldots, 4$. In terms of the $2\times2$ block decomposition (\ref{PDtilde}) we identify
\begin{equation}
\label{blocks}
\Asf=\left(\begin{array}{cc} 0&-\partial_x\\ -\partial_x&0\end{array}
\right) \qquad\Bsf^T=\left(\begin{array}{cc} 0&\wit{P}_{14}\\ 0&\wit{P}_{24}\end{array}
\right) \qquad \Csf=\left(\begin{array}{cc} 0&\wit{P}_{34}\\ -\wit{P}_{34}^T&\wit{P}_{44}\end{array}
\right)\, .
\end{equation}

The element $\wit{P}_{34}$ is given by
\begin{equation}
\wit{P}_{34}\equiv -\partial_x\cdot \left(\delta_{\mu_1}\phi_2+\delta_{\mu_2}\phi_2\right)=
-\partial_x\cdot\left(\dsl{\frac{\eta_1}{\rho_1}+\frac{\eta_2}{\rho_2}+
\frac{\eps^2}{3}\, \partial_x\cdot (\frac{\eta_1^3}{\rho_1}+\frac{\eta_2^3}{\rho_2})\cdot \partial_x}\right)\, , 
\end{equation}
since $\phi_2=\dsl{\sum_{k=1}^2\frac1{\rho_k}( \eta_k\mu_k+\frac{\eps^2}{3}(\eta_k^3\mu_{k\, x})_x)}$. 
Thus $\wit{P}_{34}$ as well as its transpose is an invertible element in the algebra of (pseudo)-differential operators on the line, so that we can apply the Dirac procedure.

By direct inspection,  the inverse of the block $\Csf$ in \eqref{blocks} is given by
\begin{equation}
\Csf^{-1}=\left(\begin{array}{cc}\medskip \wit{P}_{34}^{-1\, T}\wit{P}_{44}\wit{P}_{34}^{-1}&-(\wit{P}_{34}^{-1})^{T}\\ \wit{P}_{34}^{-1} &0\end{array}
\right)\, .
\end{equation}
Thanks to the specific forms of the blocks $\Bsf, \Bsf^T$ and $\Csf^{-1}$ the second summand in~\eqref{pdr} 
identically vanishes, since
\begin{equation}
   \Bsf^T\cdot \Csf^{-1}\cdot \Bsf= \left(\begin{array}{cc} 0&\wit{P}_{14}\\ 0&\wit{P}_{24}\end{array}
\right)\cdot \left(\begin{array}{cc} \wit{P}_{34}^{-1\, T}\wit{P}_{44}\wit{P}_{34}^{-1}&(-\wit{P}_{34}^{-1})^{T}\\\wit{P}_{34}^{-1} &0\end{array}
\right)\,\cdot \left(\begin{array}{cc} 0& 0\\ \wit{P}_{14}^T&\wit{P}_{24}^T\end{array}\right)=\left(\begin{array}{cc} 0&0\\0&0\end{array}\right)\, .
\end{equation}
This yields the   Dirac-reduced  Poisson tensor in the coordinates $(\zeta, \sigma)$ on the constrained manifold $\CC$ defined by $\{\phi_1=0,\phi_2=0\}$ as
\begin{equation}
\label{Pdirfin}
P^D=-\left(\begin{array}{cc} 0&\partial_x\\ \partial_x&0\end{array}
\right)
\end{equation}
which coincides with  the canonical Marsden-Ratiu reduced one \eqref{Pred} obtained  in Section \ref{2lay}.

\subsection{The restricted Hamiltonian}

{\color{black} To recover the Dirac reduced equations, the next task is to compute (at order $O(\epsilon^2)$) the Hamiltonian \eqref{HMCC} restricted to the manifold. To this end, we notice that using the first constraint $\phi_1$, the contribution given by $P(x,t)$ disappears and   the potential energy reduces to
\begin{equation}
    U_{MCC} = \frac{1}{2} g \left(\rho_2 - \rho_1\right)\zeta^2 \, .
\end{equation}
As usual, the non-trivial task is the computation of the  kinetic energy
\begin{equation}\label{TCC} \begin{split} 
T_{MCC}
&=\frac12\left(\frac1{\rho_1}(\eta_1\mu_1^2-\frac{\eps^2}{3}\eta_1^3\mu_{1\, x}^2)\right) +
\frac12 \left(\frac1{\rho_2}(\eta_2\mu_2^2-\frac{\eps^2}{3}\eta_2^3\mu_{2\, x}^2)\right)\,\end{split}
\end{equation}
in terms of the new variables adapted to the constraints given in \eqref{chv}.

We recall the very definition of the free variable $\sigma$, insert  $\mu_2 = \sigma + \mu_1$ in $\phi_2=0$  
and arrive at 
\begin{equation}
\left(\frac{\eta_1}{\rho_1}+\frac{\eta_2}{\rho_2}\right)\mu_1 + \frac{\epsilon^2}{3}\left(\left(\frac{\eta_1^3}{\rho_1}+\frac{\eta_2^3}{\rho_2}\right)\mu_{1 x}\right)_x = -\frac{\eta_2}{\rho_2}\sigma -\frac{\epsilon^2}{3}\left(\frac{\eta_2^3}{\rho_2}\sigma_x\right)_x\, .
\end{equation}
This relation can be inverted at  $O(\epsilon^2)$, to get 
\begin{equation}\label{sigmatomu1}
    \mu_1 = -\frac{\rho_1  \eta_2}{\psi} \sigma +\frac{\epsilon^2}{3} \frac{\rho_1}{\psi}\left((\rho_2 \eta_1^3 +\rho_1 \eta_2^3)\left(\frac{\eta_2 \sigma}{\psi}\right)_x-\frac{\eta_2^3 \sigma_x}{\rho_2}\right)_x\, ,
\end{equation}
where, again,  $\psi = \rho_1 \eta_2 +\rho_2 \eta_1$. From this relation one easily finds that
\begin{equation}\label{sigmatomu2}
    \mu_2 =\mu_1+\sigma= \frac{\rho_2 \eta_1}{\psi} \sigma +\frac{\epsilon^2}{3} \frac{\rho_1}{\psi}\left((\rho_2 \eta_1^3 +\rho_1 \eta_2^3)\left(\frac{\eta_2 \sigma}{\psi}\right)_x-\frac{\eta_2^3 \sigma_x}{\rho_2}\right)_x \,.
\end{equation}
Substituting (\ref{sigmatomu1}, \ref{sigmatomu2}) in the kinetic energy \eqref{TCC}, we obtain the following expression, 
\begin{equation}
\begin{split}
    T_{MCC} =& \frac{1}{2}\left[\frac{\eta_1\eta_2}{\psi}\sigma^2-\frac{\epsilon^2}{3} \left(\frac{\eta_1^2\eta_2^2(\rho_1\eta_1+\rho_2 \eta_2)}{\psi^2}\sigma_x^2 +\frac{\rho_1 \rho_2 (h_1+h_2)^2(\rho_2\eta_1^3 +\rho_1\eta_2^3) }{\psi^4} \zeta_x^2\sigma^2  +\right.\right.\\&\left.\left. \frac{\rho_1 \rho_2 (h_1+h_2)\eta_1 \eta_2 (\eta_1^2-\eta_2^2)}{\psi^3}\zeta_x (\sigma^2)_x \right)\right]\, ,
    \end{split}
\end{equation}
which leads to the Hamiltonian, 
\begin{equation}
\begin{split}\label{HCCrestr}
   \mathcal{H}_{MCC} =& \frac{1}{2}\int \frac{\eta_1\eta_2}{\psi}\sigma^2-\frac{\epsilon^2}{3} \left(\frac{\eta_1^2\eta_2^2(\rho_1\eta_1+\rho_2 \eta_2)}{\psi^2}\sigma_x^2 +\frac{\rho_1 \rho_2 (h_1+h_2)^2(\rho_2\eta_1^3 +\rho_1\eta_2^3) }{\psi^4} \zeta_x^2\sigma^2  +\right.\\&\left. \frac{\rho_1 \rho_2 (h_1+h_2)\eta_1 \eta_2 (\eta_1^2-\eta_2^2)}{\psi^3}\zeta_x (\sigma^2)_x \right) +  g \left(\rho_2 - \rho_1\right)\zeta^2\, d\mathbf{x} ,
    \end{split}
\end{equation}
Notice that this is, up to the rescaling by a $h^*$ factor, the Hamiltonian energy \eqref{E} of Section \ref{energ}  (and, as such,  coincides with the one found in \cite{CGK05}). 

{
{\bf Remark.} The final outcome of the result of this Section is that the Dirac Reduction of the (MCC) equations \eqref{MCCeq} and the Marsden-Ratiu reduction of the Lie-Poisson structure for a heterogeneous incompressible 2D Euler fluid do indeed coincide.
Such a coincidence was already hinted at in \cite{CGK05}, where full use was made of the  Dirichlet-to-Neumann  operator framework. By means of our geometric procedure and,  we fully brought to the light such a connection. In particular, at the $O(\eps^2)$, our explicit computations to obtain the effective Hamiltonian (\ref{HCCrestr}) do coincide with the corresponding expansion of the Dirichlet-to-Neumann operator. 
This coincidence appears very natural on physical grounds, but is less obvious
from the geometrical point of view. 
Indeed, the Marsden-Ratiu reduction recalled and described in Section \ref{2lay} starts by  identifying the submanifold $\mathcal{S}$ of $2$-layer configurations within the generic space of heterogeneous incompressible flows (with no bulk vorticity), endowed with the Benjamin linear Poisson structure~\eqref{B-pb}, and uses, as characteristic MR distribution $\mathcal{D}$ the image under the Benjamin tensor  of the one-forms vanishing on  $\mathcal{S}$. The Dirac reduction starts from a manifold which is the product of two copies of a SGN water wave model, each endowed with the ``Darboux"  tensor~\eqref{P-D}. Then  we  identificaty  the primary  constraint as the total height  of the channel and the secondary  constraint ensuing from the mass conservation laws~\eqref{appmce}. The Dirac formulation of constrained Hamiltonian systems is finally applied.  
The canonical representation~\eqref{Heqm} of the SGN equation described in Section \ref{sect: SGN} is instrumental in building this bridge, especially from the technical point of view.
}}}
\section{The large lower layer limit}\label{SecDW}
Along the same lines of  Section~\ref{sect: SGN}, we now devise a suitable scaling in which a local dispersive model can be given for "large" lower layer (i.e. deep-water) configurations (Figure \ref{Fig3}).
We reconsider the two-layer setting of section \ref{rescvar}, but use as scaling parameter  for the vertical length the upper layer asymptotic thickness $h_1$, that is, replace (\ref{xz-scale}) with
\begin{equation}\label{xz1-sc} 
x=L x^*, z=h_1\, z^*, \quad \text{with } \eps=h_1/L\, .
\end{equation}
\begin{figure}[ht!]
\centering
\includegraphics[width=13cm]{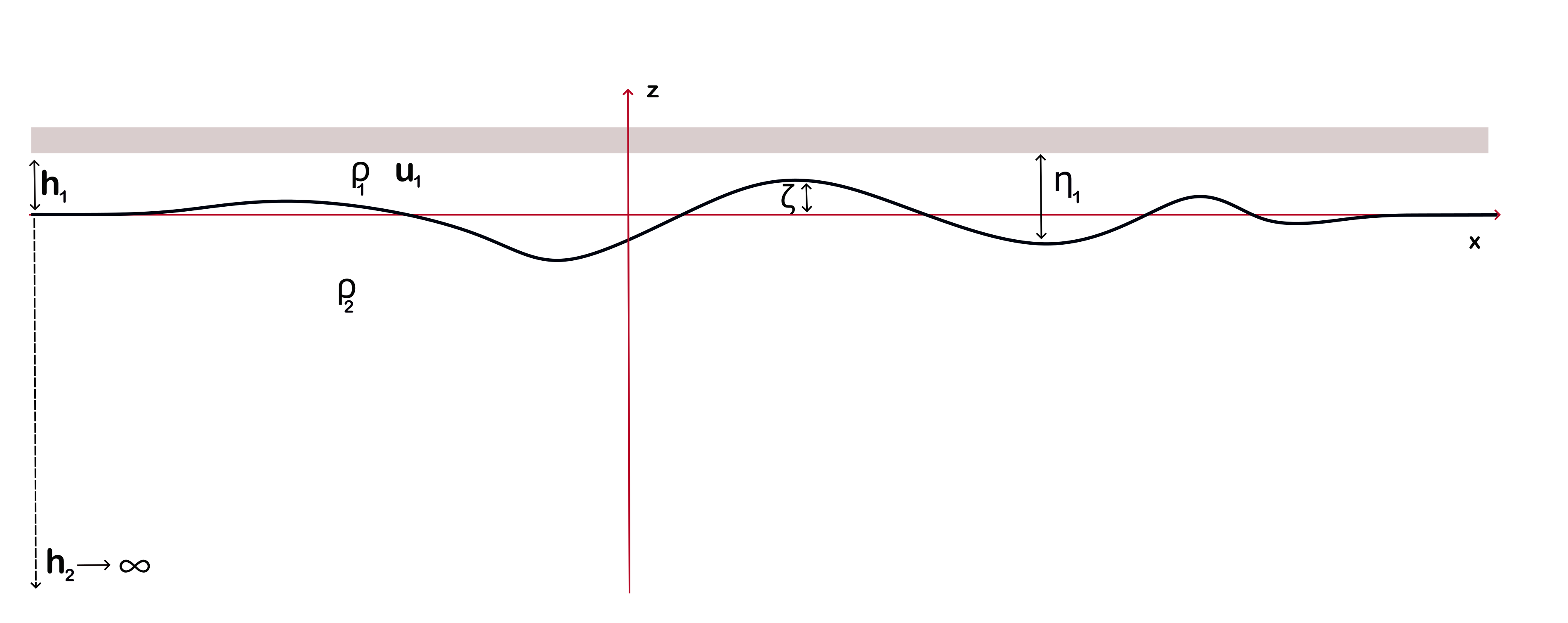}
\captionsetup{width=0.75\textwidth, font=small}
\caption{ The large lower layer ("deep water") limit. { The vertical-length scaling parameter is now the asymptotic height of the upper fluid $h_1$. The asymptotics $h_2\to\infty$ is performed. The densities $\rho_2>\rho_1>0$ are constant parameters in the scaling. As it will be shown in this Section,  the model reduces to a water wave-type model for the upper layer, the presence of the lower layer being reflected in the nonlinearity and dispersion parameters of the wave model. }}
\label{Fig3}
\end{figure}
The expression for the Hamiltonian at $O(\eps^2 )$  is  given by (\ref{E}) with $h^* = h_1$, i.e.
\begin{equation}\label{H2dw}
    \begin{split}
     H  &=  \frac{h_1 \eta_2 \eta_1}{2 \psi}\sigma^2 - \epsilon^2 \, h_1 \, \left[
\frac{\eta_2^2 \eta_1^2 \left(\rho_1 \eta_1 + \rho_2 \eta_2\right)}{6 \psi^2} \sigma_x^2 +\frac{h\, \rho_1 \rho_2 \eta_1\eta_2\left(\eta_1^2-\eta_2^2\right)}{6 \psi^3}\zeta_x (\sigma^2)_x \right. \\[10pt] & \left.   + \frac{ h^2\, \rho_1 \rho_2\left(\rho_2\eta_1^3+\rho_1 \eta_2^3\right)}{6 \psi^4}\zeta_x^2\sigma^2\right]+\frac12 h_1^2 g (\rho_2-\rho_1) \zeta^2\, ,\end{split}
\end{equation}
where we recall that $\psi=\eta_1\rho_2+\eta_2\rho_1$,  while, now,  the rescaled vertical lengths are 
\begin{equation}
\label{dw-eta's}
h=\dfrac{h_1+h_2}{h_1}=1+\dfrac{h_2}{h_1},\quad \eta_1=1-\zeta,\quad \eta_2=\dsl{\frac{h_2}{h_1}}+\zeta\, .
\end{equation} 
While the dispersionless limit of~\eqref{H2dw} is straightforward (see below),  the limit of the dispersive term is much more subtle. Indeed, it is not difficult to notice that in such an asymptotics, the dispersive term in (\ref{H2dw}) blows up linearly with $h_2$.  To circumvent this drawback, we consider the double scaling limit
\begin{equation}\label{deltascal}
    h_2 \to \infty, \quad \frac{h_1}{h_2} = O(\epsilon^a)=O\big((\frac{h_1}{L})^a\big),\quad  0<a\ll 1\, , 
\end{equation}
and denote with $\delta=\eps^{2-a}$ the new long wave expansion parameter. Even if  $\delta^2$ is not of order $\eps^4$,  $\delta^2\ll \eps^3$, and so we are entitled  to retain  only linear terms in $\delta$.

We can rewrite (\ref{H2dw}) as
\begin{equation}\label{H2dwh1h2}
    \begin{split}
     H  &=  \frac{h_1 \eta_2 \eta_1}{2 \psi}\sigma^2 - \delta \, \frac{h_1^2}{h_2} \, \left[
\frac{\eta_2^2 \eta_1^2 \left(\rho_1 \eta_1 + \rho_2 \eta_2\right)}{6 \psi^2} \sigma_x^2 +\frac{h\, \rho_1 \rho_2 \eta_1\eta_2\left(\eta_1^2-\eta_2^2\right)}{6 \psi^3}\zeta_x (\sigma^2)_x \right. \\[10pt] & \left.   + \frac{ h^2\, \rho_1 \rho_2\left(\rho_2\eta_1^3+\rho_1 \eta_2^3\right)}{6 \psi^4}\zeta_x^2\sigma^2\right]+\frac12 h_1^2 g (\rho_2-\rho_1) \zeta^2\, ,\end{split}
\end{equation}
and notice that 
since in this  asymptotics
$
\dsl{\frac{\eta_2}{\psi} \simeq \dfrac{h_2}{\psi} \to \frac1{\rho_1}},\ \text{as } h_2\to\infty$, 
we arrive to the effective "deep water" $O(\delta)$--Hamiltonian density
\begin{equation}
H_{dw}[\eta_1,\sigma]=\frac{h_1\eta_1}{2\rho_1}\sigma^2-\delta\, \frac{h_1^2\rho_2}{6\rho_1^2} 
\left((\eta_1\sigma)_x\right)^2
+\frac12 h_1^2 g (\rho_2-\rho_1) (\eta_1-h_1)^2\, . 
\end{equation}

Using as Hamiltonian variables $(\eta_1, \sigma)$, where $\eta_1=h_1-\zeta$ is the thickness of the upper layer, the Poisson tensor~\eqref{Pred} becomes
\begin{equation}
P'=\begin{pmatrix}
0&\partial_x\\
\partial_x&0
\end{pmatrix}\, , 
\end{equation}
and the ensuing equations of motion are 
\begin{equation}
\label{eqHdws-eta}
\left\{
\begin{array}{l}\medskip
\dsl{\eta_{1 \, t}-\frac{h_1}{\rho_1}(\eta_1\sigma)_x-\frac{\delta}{3}\frac{h_1^2\rho_2}{\rho_1^2}\left(\eta_1(\eta_1\sigma)_{xx}\right)_x=0}\\
\dsl{\sigma_t-\frac{h_1}{\rho_1}\sigma\sigma_x{+}g\, h_1^2(\rho_2-\rho_1)\eta_{1 \, x}-\frac{\delta}{3}\frac{h_1^2\rho_2}{\rho_1^2}\left(\sigma(\eta_1\sigma)_{xx}\right)_x=0}\,.
\end{array}
\right.
\end{equation}
{\begin{prop}
The "local deep water equations" (\ref{eqHdws-eta}) are equivalent at $O(\delta)$  to the Boussinesq-type water wave system 
\begin{equation}%\label{newdwsys}
\label{eqboufav}
\left\{
\begin{array}{l}\medskip
\dsl{\eta_{1\, t}+{h_1}(\eta_1\ou_1)_x=0}\\
\dsl{\ou_{1\, t}+h_1\ou_1\ou_{1\, x}+g\, h_1^2(\frac{\rho_2}{\rho_1}-1)\eta_{1\, x}+\frac{\delta}{3}\frac{\rho_2}{\rho_1}\left(\eta_1\right)_{xtt}=0}
\end{array}
\right.
\end{equation} 
where $\ou_1$ is the averaged upper layer velocity.
\end{prop}
{\bf Proof}. The proof can be done in a few  steps, which we  sketch. 
At first we can express the Hamiltonian variable $\sigma$ as a function of the  interface velocity of the upper layer $\wit{u}_1$ .
 The analogue of~\eqref{u2tosigma} reads
\begin{equation}
\begin{split}
       \sigma &= -\frac{\psi}{\eta_2} \wit u_1 +\epsilon^2 \frac{\rho_2}{3}\frac{\eta_1}{ \eta_2} (\eta_2^2-\eta_1^2)\wit u_{1 xx}-\\ &
        \epsilon^2  \left(\left(\frac{2}{3}\rho_2 h - (\rho_2 -\rho_1)\eta_1\right)\eta_{1 x }\wit u_{1 x}+ \frac{\rho_2 h}{3}  \left(\frac{ \eta_{1 x}^2}{\eta_2} + \eta_{1 xx}\right) \wit u_1\right)\, .
       \end{split}
\end{equation}
In the new long wave parameter $\delta$, taking the $h_2\to\infty$ limit, this reduces to 
\begin{equation}
\label{ut2tosigma}
\sigma=\dsl{-\rho_1\wit u_1+\frac{\delta}{3} h_1\rho_2\left(\eta_1\wit u_1\right)_{xx}}=\dsl{-\rho_1\ou_1+\frac{\delta}{3} h_1\rho_2\left(\eta_1\ou_1\right)_{xx}}\, .
\end{equation}
since, 
at $O(\delta) \ \ou_1=\wit{u}_1$.

After some straightforward algebra the system \eqref{eqHdws-eta} becomes at $O(\delta)$
\begin{equation}
\label{eqHdws-eta2}
\left\{
\begin{array}{l}\medskip 
\dsl{\eta_{1\, t}+{h_1}(\eta_1\ou_1)_x=0}\\
\dsl{\ou_{1\, t}+h_1\ou_1\ou_{1\, x}+g\, h_1^2(\frac{\rho_2}{\rho_1}-1)\eta_{1\, x}-\frac{\delta}{3}h_1\frac{\rho_2}{\rho_1}\left(\eta_1\ou_1\right)_{xxt}=0}
\end{array}
\right.
\end{equation}
Finally, by using the ``mass conservation" equation $\eta_{1\, t}+{h_1}(\eta_1\ou_1)_x=0$, system  \eqref{eqHdws-eta2} can be traded for the Boussinesq-favoured system~\eqref{eqboufav}.}
\hfill$\blacksquare$\medskip\\ \\ \\
{\bf Remarks.} 
\begin{enumerate}
   \item 
   The map \eqref{ut2tosigma}, besides identifying the model as one more signal of the universality of Boussinesq models, serves also as a regularizing map. Indeed, as it often happens in such effective models, system \eqref{eqboufav} is linearly well-posed, contrary to  the purely evolutive Hamiltonian system \eqref{eqHdws-eta}.

\item In the large $h_2$ asymptotics the dispersionless Hamiltonian reduces to 
\begin{equation}
\label{Hdwdisp}
H_{dw}=\frac{h_1\eta_1}{2\rho_1}\sigma^2+\frac12 h_1^2 g (\rho_2-\rho_1) \zeta^2\, , 
\end{equation}
engendering  equations of motion equivalent to those of (\cite{CaCh16,CC99}, equations (4.17-18) ), i.e,
\begin{equation}
\label{eqBoudl}
\left\{
\begin{array}{l}\medskip
\dsl{\eta_{1\, t}+h_1(\eta_1\ou_1)_x=0}\\
\dsl{\ou_{1\, t}+h_1\ou_1\ou_{1\, x}+h_1^2 g(\frac{\rho_2}{\rho_1}-1)\eta_{1\, x}=0}\, .
\end{array}
\right.
\end{equation}
\item  {
Our asymptotic expansion is no longer valid at $O(\eps^2)$, but rather at the first order in the long-wave parameter $\delta =\eps^{2-a}$ with $0<a\ll1$. 
The difference of our setting with  that of the standard approach of \cite{B67,CaCh16, CC99}, where the asymptotic model for a stratified fluid with a large lower layer is a Benjamin-Ono type system,  relies on the relative scaling between $h_2$ and $L$. Indeed, in   \cite{B67,CaCh16, CC99} the asymptotic regime consists in  ${h_2}/{L}=O(1)$. We use the scaling ${h_2}/{L}=O(\eps^{1-a})$,  implying that  perturbations of the quiescent state can somehow be considered as  waves in  a shallow fluid  with respect to the lower layer also. This is the physical explanation for the appearance of a Boussinesq type local system as the effective model.}

\end{enumerate}
\section{Conclusions}
This paper is devoted to framing effective wave models in one space dimension within the Hamiltonian setting of $1+1$-dimensional PDEs. Our main aim is to show that the Hamiltonian structure of effective models, associated with sharply stratified field configurations, indeed originates from the parent full Hamiltonian structure of the $2+1$-dimensional Euler system. To this end, we found it particularly convenient to start with the Benjamin representation of such a structure~\cite{Ben86}.

Thanks to our reduction scheme, the resulting reduced Poisson structure turns out to be independent of the physical parameters entering the sharp stratification, namely the asymptotic heights $h_1, h_2$ and the densities $\rho_1,\rho_2$ of the fluid's "phases". Thus, different interesting effective long-wave models associated with sharply stratified configurations can be obtained simply by fine-tuning parameters and asymptotic expansions.

{At first, in Section~\ref{energ}, we find the fully non-linear 2-layer model extending the dispersionless, { Weakly and Mildly Non-Linear} approach of ~\cite{CFO17,CFOPT23} to cover such a case. We make here the choice to scale vertical lengths by  a generic parameter $h^*$ that we specialize to attain different physical values according to the problem we tackle.  

At first, setting $h^*=h$, the total heights of the channel,  our findings match the ones in~\cite{CGK05} without explicitly relying on the Dirichlet-to-Neumann formalism.  

We then  choose $h^*=h_2$ and,
applying a natural double-scaling procedure, we  directly obtained a canonical form of the SGN equations.
Exploiting this we finally provide the fully nonlinear Miyata Choi-Camassa model with a Hamiltonian formulation. We then perform an infinite-dimensional version of the celebrated Dirac reduction procedure to obtain the the fully reduced MCC model (i.e., the unconstrained model in two independent variables) in the Hamiltonian formalism. Closing this circle of ideas, we show that this coincides with the one obtained in Section~\ref{energ} by means of the Marsden-Ratiu-like reduction on the two-layered configuration space.

In the final chapter we reconsidered the general formulas of Section~\ref{energ} choosing as vertical length scale $h^*=h_1$ the asymptotic height of the {\em upper} fluid and let
the width of the heavier lower fluid grow indefinitely. We have shown that with a suitable double scaling limit this procedure  leads to a local model for "deep-water" waves, which can be finally identified with a  Boussinesq-type system, with nonlinearity and dispersion parameters dictated by the density ratio ${\rho_1}/{\rho_2}$.

 }

\paragraph{Authors Contribution Declaration:} G.F. and E.S.  have devised and  developed the ideas of the present paper, performed the necessary computations, and reviewed the manuscript. E.S. provided Figures 1, 2, 3, 4.

\subsection*{Acknowledgments}

This work has received funding by the Italian PRIN 2022 (2022TEB52W) - PE1 - project {\em The charm of integrability: from nonlinear waves to random matrices}. We also gratefully acknowledge the auspices of the GNFM Section of INdAM, under which part of this work was carried out, and the support of the project  MMNLP (Mathematical Methods in Non Linear Physics) of the Italian INFN. The authors are  indebted to R. Camassa, G. Ortenzi and M. Pedroni for sharing their insights about the theory of stratified fluids in a long-standing collaborative effort, of which the present paper can be seen as an outgrowth.


\begin{thebibliography}{99}
\bibitem{B67}{Benjamin, T. B.}, \emph{Internal waves of permanent form in fluids of great depth}, 
 J. Fluid Mech. \textbf{29} (1967), 559--592.
  \bibitem{Ben86} {Benjamin, T. B.}, \emph{On the Boussinesq model for two-dimensional wave motions in heterogeneous fluids}, 
 J. Fluid Mech. \textbf{165} (1986), 445--474.
\bibitem{BB97}  
Benjamin, T. B., Bridges, T. B., {\em Reappraisal of the Kelvin-Helmholtz problem. Part 1.
Hamiltonian structure}, J. Fluid Mech. {\bf 333} (1997), 301--325.

\bibitem{CCFOP12} 
{Camassa, R., Chen, S.,  Falqui, G., Ortenzi, G., Pedroni, M.}, \emph{An inertia `paradox' for incompressible stratified Euler fluids},
 J. Fluid Mech. \textbf{695} (2012), 330--340.

 \bibitem{CFOP14}
 {Camassa, R.,  Falqui, G., Ortenzi, G., Pedroni, M.}, {\em On variational formulations and conservation laws for incompressible $2D$ Euler fluids}, Journal of Physics: Conference Series {\bf 482} (2014) 012006
\bibitem{CFO17} 
Camassa, R., Falqui, G., Ortenzi, G.,  {\it Two-layer interfacial flows beyond the Boussinesq approximation: a Hamiltonian approach},
Nonlinearity {\bf 30}  
(2017), 466--491.
\bibitem{CFOPS26} Camassa, R., Falqui, G., Ortenzi, G., Pedroni, M., Sforza, E. {\em Two-layer sharply stratified Euler fluids in three dimensions: a Hamiltonian setting}, arXiv:2604.22622, to appear in  Nonlinearity.
\bibitem{CFOPT23} Camassa, R., Falqui, G., Ortenzi, G., Pedroni, M. Vu Ho, TT, {\em Simple two-layer dispersive models in the Hamiltonian reduction formalism,} Nonlinearity \textbf{36} (2023), 4523---4552.
\bibitem{CHH94}Camassa, R., Holm, D. D.,  Hyman, J. M., {\em A New Integrable Shallow Water Equation} Adv. Appl. Mech. \textbf{31} (1994), 1---33.

\bibitem{CaCh16} 
Choi, W., Camassa, R., \emph{Weakly nonlinear internal waves in a two-fluid system}, J. Fluid Mech. {\bf 313} (1996), 83--103. 
\bibitem{CC99} 
Choi, W., Camassa, R., \emph{Fully nonlinear internal waves in a two-fluid system},  J. Fluid Mech. \textbf{396} (1999), 1--36.
\bibitem{cbj}
{Choi, W., Barros, R., Jo, T.-C., {\em 
A regularized model for strongly nonlinear internal solitary waves}, {J. Fluid Mech.} 
{\bf 629}  (2009),  73--85.}

\bibitem{Chumaetal09} 
Chumakova, L., Menzaque, F. E., Milewski, P. A., Rosales, R. R., Tabak, E. G., Turner, C. V., {\em Stability properties and nonlinear mappings of two and three-layer stratified flows}, Stud. Appl. Math. {\bf 122} (2009), 123--137.

\bibitem{CI19} Constantin, A.,  Ivanov, R. I., 
{\em Equatorial wave-current interactions}, 
Comm. Math. Phys. {\bf 370} (2019), 1–48.
\bibitem{CGK05}  
Craig, W., Guyenne, P., Kalisch, H., {\em Hamiltonian long-wave expansions for free surfaces and interfaces}, Comm. Pure Appl. Math. {\bf 58} (2005), 1587--1641.
\bibitem{CS93}  
Craig, W., Sulem, C., {\em Numerical simulation of gravity waves}, J. Comput. Phys. {\bf 108} (1993), 73--83.
\bibitem{Dirac}
Dirac, P. A. M., {\em Generalized Hamiltonian dynamics},  Canad. J. Math. {\bf 2} (1950), 129--148.
\bibitem{Du16}
Duch\^{e}ne, V., Israwi, S., Talhouk, R., {\em A new class of
two-layer Green-Naghdi systems with improved frequency dispersion},
Stud. Appl. Math. {\bf 137} (2016), 356--415.
\bibitem{G-BMR12}   Gay-Balmaz, F., Marsden, J. E., Ratiu,  T. S., {\em Reduced Variational Formulations in Free Boundary Continuum Mechanics}, 
J Nonlinear Sci {\bf 22}  (2012), 463–497.
{
\bibitem{GN76} Green,  A. E., Naghdi, P. M., {\em A derivation of equations for wave propagation in water of variable depth}, J. Fluid Mech. {\bf 78} (1976), 237-246.}

\bibitem{HMR98} Holm, D. D., Marsden, J. E., Ratiu,  T. S.,{\em The Euler-Poincar\'e Equations and Semidirect Products with Applications to Continuum Theories},  Adv. Math. {\bf 137} (1998), 1-81.
{
\bibitem{LB09}
Lannes, D., Bonneton, P., {\em Derivation of asymptotic two-dimensional time-dependent equations for surface water wave propagation}, Phys. Fluids {\bf 21} (2009), 016601.}
 \bibitem{Lin63} Lin,  C. C.,  {\em Hydrodynamics of Liquid Helium II}, Proc. Int. School of Physics E. Fermi, New York, Academic
Press (1963) Reprinted in 1987 Selected papers of C C Lin, Singapore, World Scientific.
\bibitem{MR86} 
 Marsden, J. E., Ratiu, T. S.,  {\em Reduction of Poisson manifolds}, Lett. Math. Phys. {\bf 11} (1986), 161--169.
 \bibitem{Ma15} Matsuno, Y., {\em Hamiltonian formulation of the extended Green–Naghdi equations},  Physica D {\bf 301–302} (2015), 1–7
{\bibitem{MS85}
Miles, J., Salmon, R.,{\em Weakly dispersive nonlinear gravity waves}, J. Fluid Mech. (1985), {\bf 157}  519-531.}
 {
 \bibitem{M88} Miyata, M., {\em Long Internal Waves of Large Amplitude}, in: Nonlinear Water Waves. International Union of Theoretical and Applied Mechanics (Horikawa, K., Maruo, H., eds), Springer (Berlin, Heidelberg), 1988.}
   \bibitem{OVS79} 
 Ovsyannikov, L. V., \emph{Two-layer ``shallow water" model}, J. Appl. Mech. Tech. Phys. \textbf{20} (1979), 127--135.
 
 \bibitem{PCH} 
Percival, J. R., Cotter, C. J., Holm, D. D.,
{\em A Euler--Poincar\'e framework for the multilayer Green--Nagdhi equations},
{J. Phys. A}  {\bf 41} (2008), 344018. 
\bibitem{Ser56} Serre, F.,  {\em Contribution \`a l’\'etude des \'ecoulements permanents et variables dans les canaux}, La Houille Blanche {\bf 8} (1953), 374–388.
\bibitem{SG69} Su, C.H., Gardner, C. S., {\em KDV Equation and Genaralizations. Part III. Derivation of Korteweg-de Vries Equation and Burgers Equation}, J. Math. Phys. {\bf 10} (1969) 536–539.
\bibitem{Wh2000} 
 Whitham, G. B., Linear and Nonlinear Waves, Wiley \& Sons (New York), 2000.
 \bibitem{Wu81} 
 Wu, T. Y., {\em Long waves in ocean and coastal waters}, J. of Eng. Mech., \textbf{107} (1981), 501--522.
 \bibitem{Wu98} 
   Wu, T. Y., {\em Nonlinear waves and solitons in water}, Physica D {\bf 123} (1998), 48--63.
 \bibitem{Wu2000} 
 Wu, T. Y., {\em On Modeling Unsteady Fully Nonlinear Dispersive Interfacial Waves}, in: Fluid Dynamics at Interfaces (Shyy, W., and Narayanan, R., eds.), Cambridge Univ. Press, 2000.
 \bibitem{Zak68} 
{Zakharov, V. E.,} \emph{Stability of periodic waves of finite amplitude on the surface of a deep fluid},  Zh. Prikl. Mekh. Tekh. Fiz. \textbf{9} (1968), 86--94.
\bibitem{ZK97}
{Zakharov, V. E., Kuznetsov E. A.}\emph{Hamiltonian formalism for nonlinear waves,} Phys. Usp. {\bf  40} (1997), 1087–1116
\bibitem{Z85} 
{Zakharov, V. E., Musher, S. L., Rubenchik, A. M.}, \emph{Hamiltonian approach to the description  of non-linear plasma phenomena},  Phys. Rep.  \textbf{129} (1985), 285--366.

\end{thebibliography}
\end{document}